\documentclass[aps,prl,reprint,showpacs,superscriptaddress]{revtex4-1}

\usepackage{graphicx}
\usepackage{amsmath}
\usepackage{amssymb}
\usepackage{color}
\usepackage[normalem]{ulem}

\begin{document}

\title{Dephasing due to nuclear spins in large-amplitude electric dipole spin resonance}

\author{Stefano Chesi}
\email[]{stefano.chesi@csrc.ac.cn}
\affiliation{Beijing Computational Science Research Center, Beijing 100084, China}

\author{Li-Ping Yang}
\affiliation{Beijing Computational Science Research Center, Beijing 100084, China}

\author{Daniel Loss}
\affiliation{Department of Physics, University of Basel, Klingelbergstrasse 82, 4056 Basel, Switzerland}
\affiliation{CEMS, RIKEN, Wako, Saitama 351-0198, Japan}

\date{\today}

\begin{abstract}
We have analyzed effects of the hyperfine interaction on electric dipole spin resonance when the amplitude of the quantum-dot motion becomes comparable or larger than the quantum dot's size. Away from the well known small-drive regime, the important role played by transverse nuclear fluctuations  leads to a gaussian decay with characteristic dependence on drive strength and detuning.  A characterization of spin-flip gate fidelity, in the presence of such additional drive-dependent dephasing, shows that vanishingly small errors can still be achieved at sufficiently large amplitudes. Based on our theory, we analyze recent electric-dipole spin resonance experiments relying on spin-orbit interactions or the slanting field of a micromagnet. We find that such experiments are already in a regime with significant effects of transverse nuclear fluctuations and the form of decay of the Rabi oscillations can be reproduced well by our theory.
\end{abstract}

\pacs{85.35.Be,75.75.-c,76.30.-v,03.65.Yz}

\maketitle

\paragraph{Introduction.} 
The interest in coherent manipulation of single electron spins has stimulated intense research efforts, leading to a great degree of control in a variety of nanostructures \cite{Hanson2007,Awschalom2013}. For electrons in quantum dots, electron spin resonance (ESR) was first demonstrated in Ref.~\cite{Koppens2006}. However, full electric control of local spins might be a better strategy for complex architectures of many quantum dots, envisioned to realize quantum information processing \cite{Loss1998}. Thus, electric dipole spin resonance (EDSR) was developed relying on either spin-orbit couplings \cite{Golovach2006,Nowack2007} or the inhomogeneous magnetic field induced by a micromagnet \cite{Tokura2006,Pioro2008}. The effectiveness of EDSR is highlighted by recent experiments, which could demonstrate Rabi oscillations with frequencies larger than 100~MHz for both approaches \cite{vdBerg2013,Yoneda2014}. To further improve the performance of such spin manipulation schemes, it is important to characterize relevant dephasing mechanisms, and especially those which might become dominant at strong electric drive. In fact, as it will become clear in the following, a sufficiently strong drive is able to induce significant and yet unexplored modifications on
how typical dephasing sources affect EDSR. For this reason, while representing the main limitation for accurate spin manipulation, dephasing is still poorly understood in large-amplitude regime of EDSR (i.e., when the amplitude of motion becomes comparable to the quantum dot's size). 

In this work we will focus on hyperfine interactions, which are well known to play an important role in the electron spin dynamics of quantum dots. In particular, the ESR dephasing was successfully interpreted in terms of a static Overhauser field, with a variance of a few mT in GaAs \cite{Koppens2006}. The resulting power-law decay and a universal $\pi/4$ phase shift of the Rabi oscillations were accurately verified \cite{Koppens2007}, confirming the predominance of nuclear spins over other sources of dephasing. While EDSR experiments were also generally interpreted assuming a power-law decay, the expected $t^{-1/2}$ dependence is violated at the larger values of the drive \cite{NadjPerge2010,vdBerg2013,Yoneda2014}. Especially, Ref.~\cite{Yoneda2014} has demonstrated striking deviations from the ESR behavior, including a crossover from power-law to gaussian decay.  
It is also known that the electron motion, as well as the presence of the drive, can have substantial effects on spin dynamics and decoherence  \cite{Laird2007,Rashba2008,Palyi2014,Arrondo2013,Jing2014}. These considerations motivate us to provide here a detailed characterization of EDSR dephasing induced by the hyperfine interaction, paying special attention to the large-amplitude regime. As a main objective behind trying to achieve faster Rabi frequencies is to decrease operation errors, we also establish the limitations on spin-flip gate fidelity imposed on EDSR by the hyperfine interaction. Finally, we compare our theory with EDSR experiments which, as we will discuss, have very recently entered the large-amplitude regime. 

\paragraph{Model.}  EDSR is induced by a driven periodic displacement of the quantum dot $\vec{R}(t)=\hat{\bf e}_x  \, \delta R \sin \omega t$, which we take conventionally along $x$. For the time-dependent wavefunction $\psi(\vec{r}-\vec{R}(t))$, we assume harmonic confinement along the direction of motion (which is applicable to both nanowire and lateral quantum dots):
\begin{equation}\label{psi}
\left\vert \psi (\vec{r})\right\vert ^{2}=|\varphi (y,z)|^{2}\frac{1}{\sqrt{\pi}
\delta x}e^{-x^{2}/\delta x^{2}}.
\end{equation}
The spin dynamics can be described with the following Hamiltonian:
\begin{equation}\label{H}
H=\frac{\epsilon_z}{2}\sigma _{z}+\frac{\vec{b}\cdot \vec{\sigma}}{2}  \sin
\omega t  +\sum_{i}\frac{A_i}{n_0}\left\vert \psi (\vec{r}_{i}-\vec{R}(t))\right\vert ^{2}\vec{\sigma}\cdot 
\vec{I}_{i},
\end{equation}
where the first term is the electron Zeeman coupling, with $\epsilon_z=g \mu_B B$ and $\vec{\sigma}$ the Pauli matrices. The second term is the drive, for which we can generally assume $b \propto \delta R$ while other features (e.g., the direction of $\vec b$) depend on specific details of the spin-orbit coupling or magnetic gradient. The last term in Eq.~(\ref{H}) is the Fermi contact hyperfine interaction, where  $n_0$ is the nuclear density. $\vec{I}_i$ is the spin operator of nucleus $i$, with position $\vec{r}_i$ and coupling $A_i$. The periodic time-dependence of Eq.~(\ref{H}) is characterized by Fourier components
$\psi _{m}(\vec{r})=\frac{\omega }{2\pi }\int_{0}^{2\pi /\omega
}\vert \psi (\vec{r}-\vec{R}(t))\vert ^{2}e^{-im\omega t}dt$,
of which only the static ($m=0$) and resonant ($m=\pm 1$) ones are of interest here. In fact, in a frame rotating at frequency $\omega \simeq \epsilon_z/\hbar$ and neglecting fast oscillating terms, the transformed Hamiltonian $H'$ reads:
\begin{eqnarray}\label{Hrot}
H'  \simeq  && \,\frac{\epsilon_z-\hbar\omega}{2}\sigma _{z}-\frac{1}{4} \left( \vec{b} \times 
\vec{\sigma}\right)_z \nonumber\\
&&  +\sum_{i}\frac{A_{i}}{n_0}\left[\psi _{0}(\vec{r}_{i})\sigma _{z}I_{i,z}  +i\psi _{1}(\vec{r}_{i})\left( \vec{\sigma} \times \vec{I}_i \right)_z \right],
\end{eqnarray}%
where longitudinal/transverse fluctuations are controlled by $\psi _{0}(\vec{r})$ and $\psi _{1}(\vec{r})$, respectively. Without loss of generality, we restrict ourselves to the case $b_x=b_z=0$ \cite{remark_by_bz}:
\begin{equation}\label{Hsimple}
H'= \frac{\Delta}{2}\sigma _{z}+\frac{b}{4} \sigma_x + \frac12\vec{h} \cdot \vec{\sigma},
\end{equation}
where $\Delta = \epsilon_z-\hbar\omega $ is the detuning and $\vec h$ is defined by the second line of Eq.~(\ref{Hrot}). 

\begin{figure}
\begin{center}
\includegraphics[height=0.21\textwidth]{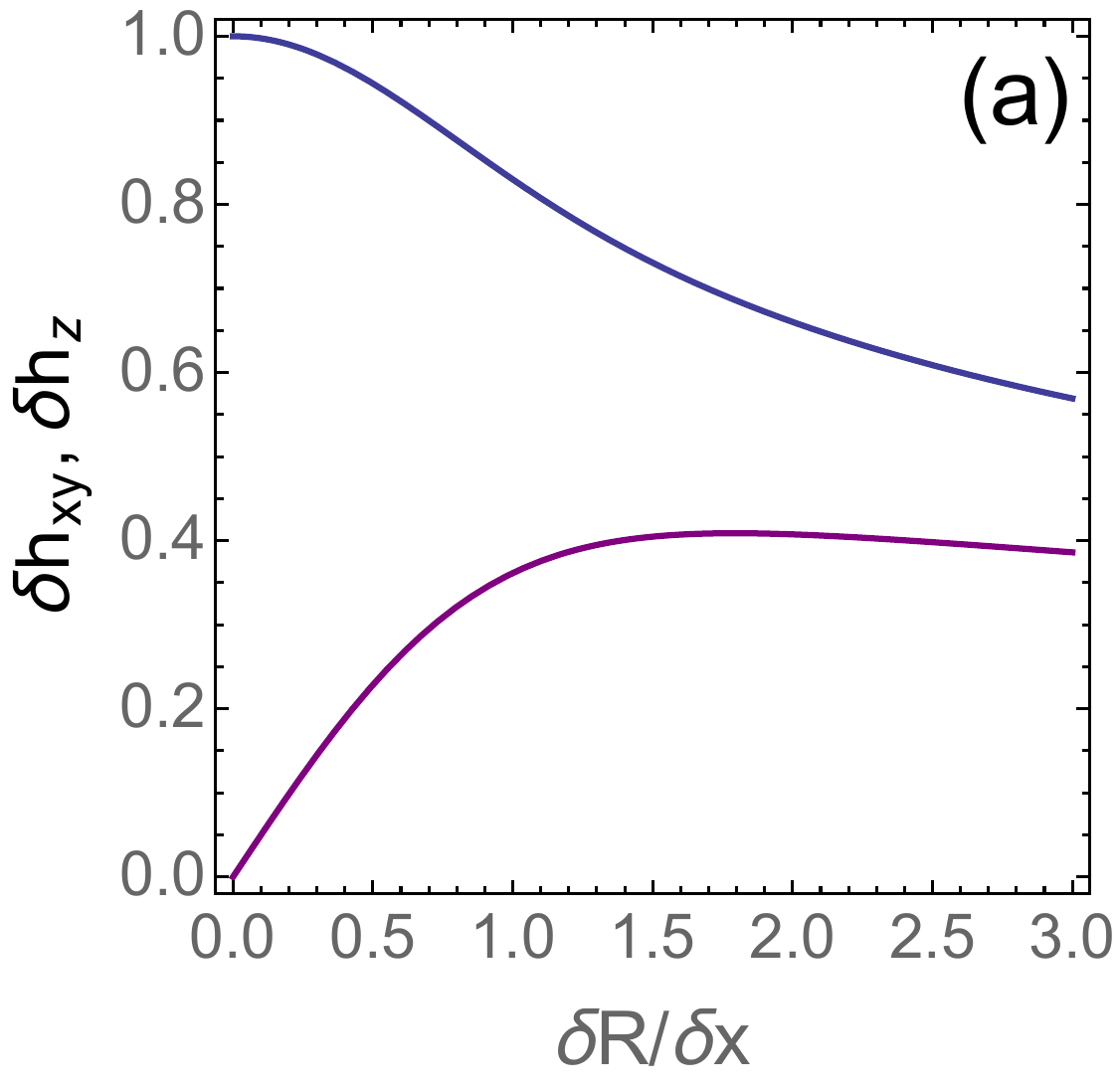}
\includegraphics[height=0.21\textwidth]{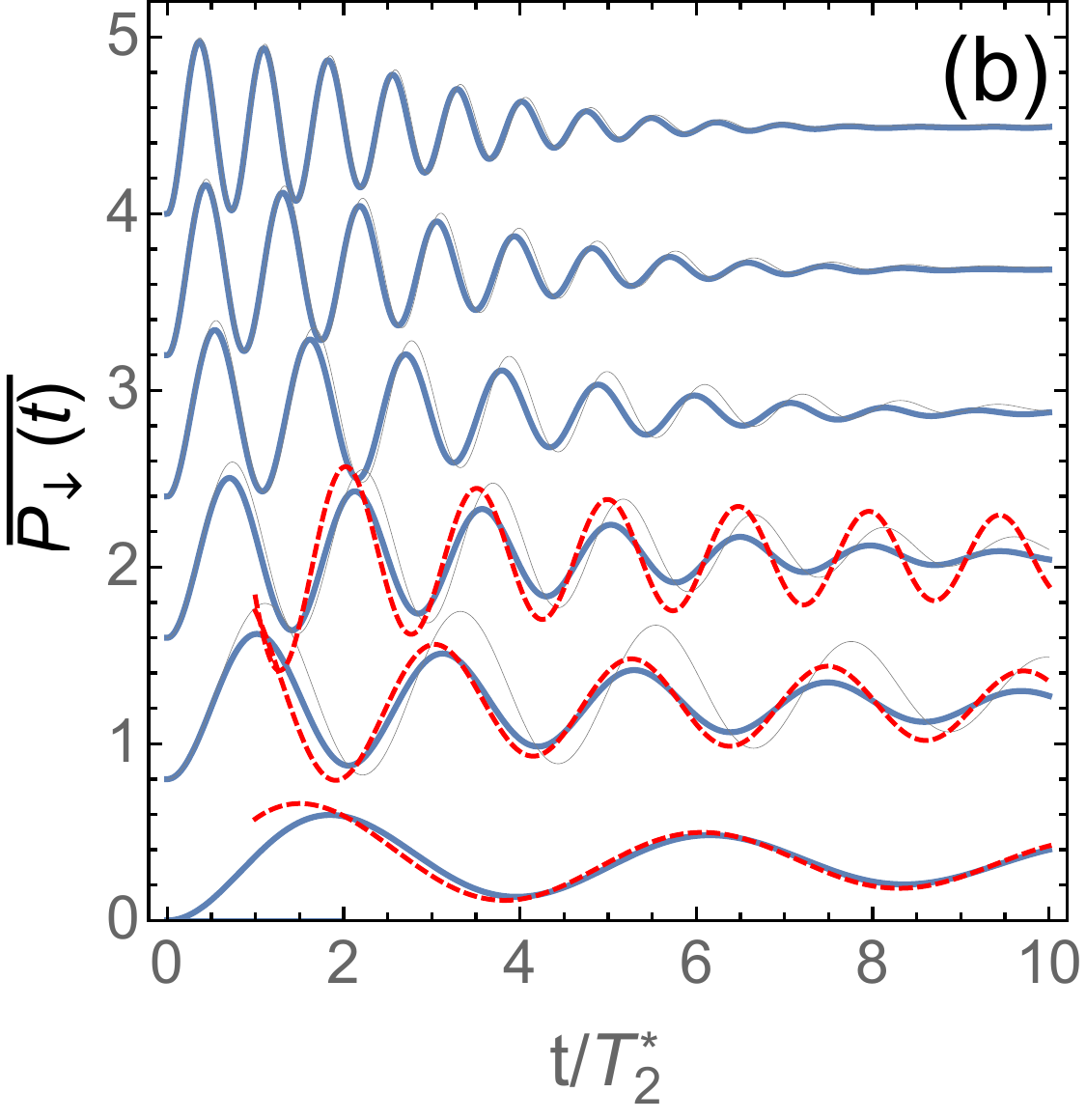}
\includegraphics[height=0.21\textwidth]{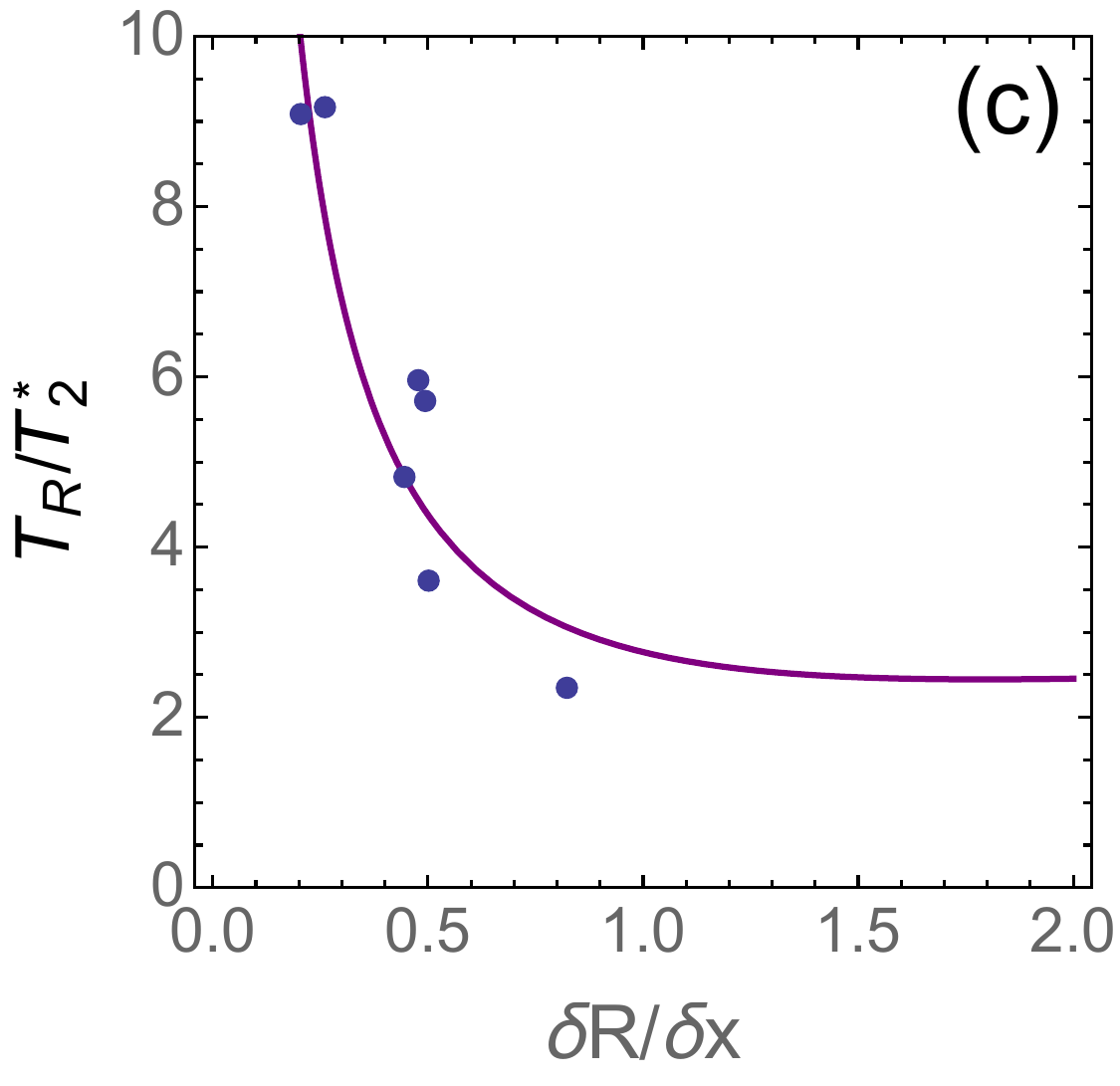}
\includegraphics[height=0.21\textwidth]{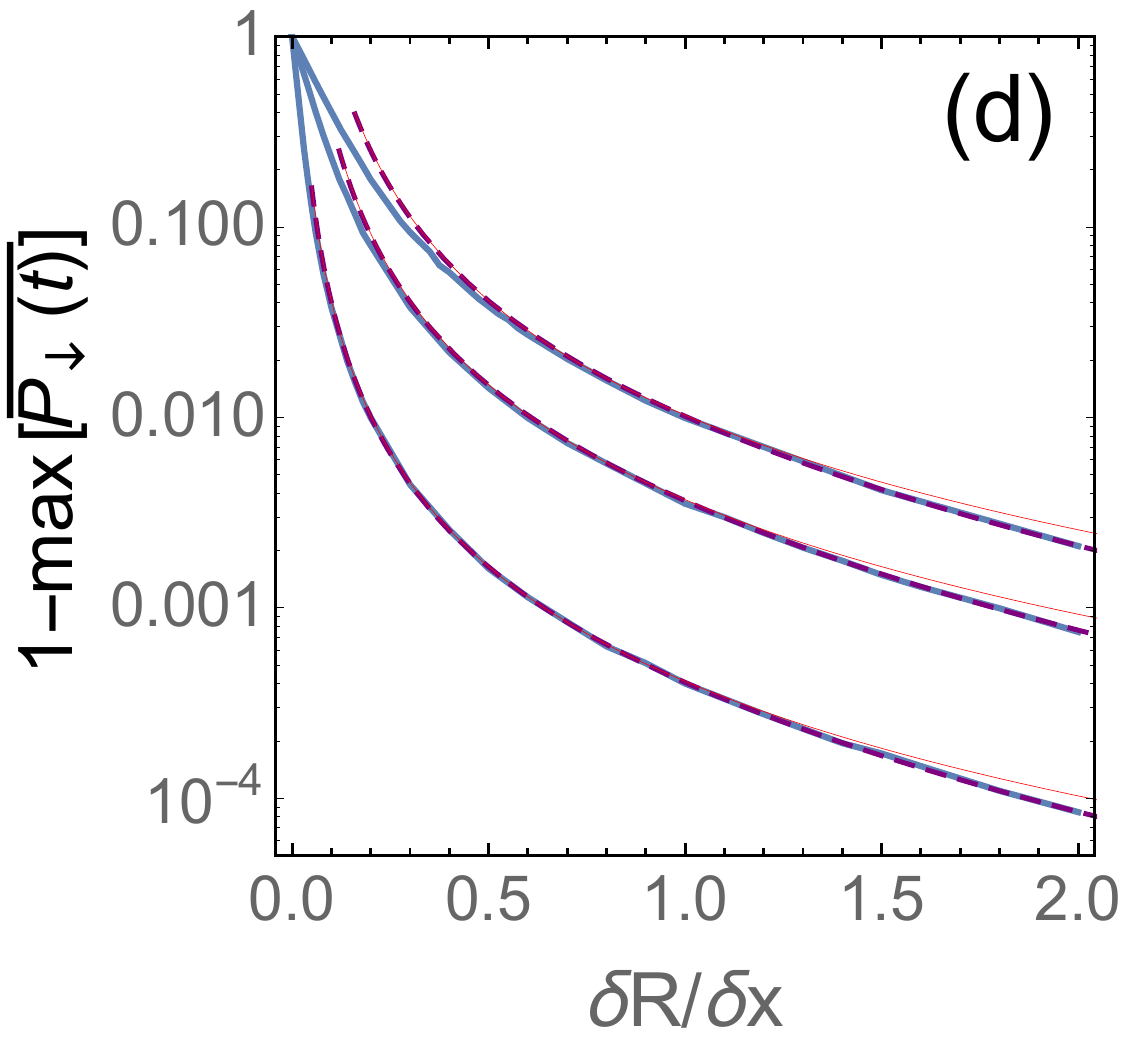}
\caption{(a): Plot of Eqs.~(\ref{hz}) and~(\ref{hxy}), which characterize the longitudinal/transverse nuclear fluctuations (upper/lower curve, respectively). (b): Numerical result for the Rabi oscillations $\overline{P_\downarrow(t)}$, averaged over nuclear fluctuations (blue thick curves). We used $\eta =0.05$, $\Delta=0$, and $\delta R/\delta x=0.1,0.2,... 0.6$ (bottom to top). For clarity, the curves are shifted vertically. The red dashed curves are the asymptotic power-law decay of Ref.~\cite{Koppens2007}, valid when $\delta R/\delta x \lesssim \sqrt{2 \eta} = 0.3 $. The thin grey curves are from Eq.~(\ref{P_detuning}). (c): Plot of Eq.~(\ref{TR0}). Dots are decay times from Ref.~\cite{Yoneda2014}, rescaled using $T_2^* \simeq  9$~ns and $\eta\simeq 0.09$. (d): Error in realizing a $\pi$-rotation, plotted as function of $\delta R/\delta x$ at $\eta=0.01, 0.03, 0.05 $ (bottom to top). Numerical results (thick solid) are compared to the asymptotic result (thick dashed) and the upper bound (thin red) given in Eq.~(\ref{infidelity}) \cite{comment_fidelity}. The thin red lines also practically coincide with the ESR result, as seen by taking $\delta h_z =1$ and $\delta h_{xy}=0$ in Eq. (\ref{infidelity}).\label{fig:1}}
\end{center}
\end{figure}

\paragraph{Nuclear fluctuations.} On the relatively short time scales of the EDSR experiments, it is appropriate to describe $\vec h$ with a static random classical magnetic field. In the lab frame and for infinite-temperature nuclear spins, the variance of the Overhauser field is given by
$
\sigma ^{2}=\sum_{i} (2A_{i}/n_0)^{2}|\psi (\vec{r_i})|^{4} I_{i}(I_i+1)/3.
$
However, $\vec h$ is for a reference frame moving with the dot and rotating at frequency $\omega$. As a consequence, its statistical properties differ from the ones in the lab frame. We still have $\langle \vec{h} \rangle =0 $, but Eq.~(\ref{Hrot}) implies that $\langle h_z^2\rangle$, $\langle h_{x,y}^2\rangle$ have an interesting dependence on the strength of the drive. For finite $\delta R$ and $\left\vert \psi (\vec{r})\right\vert ^{2}$ as in Eq.~(\ref{psi}) we can evaluate $\left\langle h_{z}^{2}\right\rangle $, $\left\langle h_{x,y}^{2}\right\rangle $ as follows, in terms of Hypergeometric functions:
\begin{eqnarray}
&& \delta h_z = \frac{\sqrt{\left\langle h_{z}^{2}\right\rangle }}{\sigma } =\sqrt{\left. _{p}F_{q}(%
\frac{1}{2},\frac{1}{2};1,1;-\frac{2\delta R^{2}}{\delta x^{2}})\right. }, \label{hz} \\
&& \delta h_{xy} = \frac{\sqrt{\left\langle h_{x,y}^{2}\right\rangle }}{\sigma } =\frac12 \frac{\delta
R}{\delta x} \sqrt{\left. _{p}F_{q}(\frac{3}{2},\frac{3}{2};2,3;-\frac{%
2\delta R^{2}}{\delta x^{2}})\right.} .    ~~~\label{hxy} 
\end{eqnarray}%
In the above formulas, the only dependence is on $\delta R/\delta x$, i.e., the amplitude of motion relative to the width of the electron wavefunction. $\delta h_z$ and $\delta h_{xy}$ are plotted in Fig.~\ref{fig:1}(a), showing that for $\delta R \to 0$ only the longitudinal fluctuations survive. Therefore, in this limit one recovers the same behaviour of ESR. At finite $\delta R/\delta x$, the value of $\delta h_z$ is a decreasing function of amplitude while the transverse fluctuations become non-zero and have a non-monotonic dependence on $\delta R/\delta x$. Such transverse fluctuations can serve at $b=0$ as a driving term \cite{Laird2007}, and in this context were previously discussed through an expansion at small $\delta R$ \cite{Rashba2008} or numerical evaluation \cite{Palyi2014}. However, both scenarios of EDSR (i.e., based on a micromagnet or spin-orbit coupling) are in a physical regime distinct from Refs.~\cite{Laird2007,Rashba2008,Palyi2014}, because $\delta h_{xy}$ is typically much smaller than the drive $b$. To see this, we notice that Eq.~(\ref{hxy}) implies $\delta h_{xy} < \frac12 \delta R/\delta x$, thus:
\begin{equation}\label{xy_vs_b}
\frac{h_{x,y}}{b} \sim \frac{\sigma \delta h_{xy}}{b}   < \frac{\delta R/\delta x}{2b/\sigma}  = \frac{\eta}{2} \ll 1.
\end{equation}
In Eq.~(\ref{xy_vs_b}) we defined the useful parameter $\eta=\sigma\delta R /b\delta x$. $\eta$ is approximately constant (since $b\propto \delta R$) and is typically small, according to our later estimates. Therefore, transverse nuclear fluctuations provide an additional dephasing mechanism which becomes progressively more important, until the maximum effect is reached at $\delta R /\delta x \simeq 1.8$. We will discuss how the effect of $\delta h_{xy}$ becomes dominant over $\delta h_{z}$  in a regime of sufficiently strong EDSR drive, which was already realized by recent experiments \cite{Yoneda2014}.

\paragraph{Rabi oscillations.} We now use Eqs.~(\ref{hz}) and (\ref{hxy}) to perform a gaussian average $\overline{P_\downarrow(t)}$ with respect to $\vec{h}$ of the spin-flip probability $P_\downarrow(t)$:
\begin{eqnarray}\label{Rabi_no_noise}
P_\downarrow(t && ) = \frac{(b/2+h_x)^2+h_y^2}{(b/2+h_x)^2+h_y^2+(\Delta+h_z)^2} \nonumber \\
&& \times \sin^2\left(\frac{t}{2 \hbar}\sqrt{\left(b/2+h_x\right)^2+h_y^2+(\Delta+h_z)^2} \right).\qquad
\end{eqnarray}
Although we cannot provide a general closed-form result, several relevant features can be explicitly characterized. In particular, at sufficiently large drive and detuning we can neglect the components of $\vec{h}$ perpendicular to $(b/2) \hat{\bf e}_x +\Delta \hat{\bf e}_z $ [see Eq.~(\ref{Hsimple})], to obtain:
\begin{equation}\label{P_detuning}
\overline{P_\downarrow (t)} \simeq \frac{b^2/2}{{b^2+4\Delta^2}} \left[ 1-e^{-\left(t/T_R \right)^2} \cos\left(\frac{t}{\hbar}\sqrt{b^2/4+\Delta^2} \right)\right].
\end{equation}
The Rabi decay time is:
\begin{equation}\label{TR0general}
T_R(\Delta) = \left( \frac{T_2^* }{\delta h_{xy}} \right) \times \sqrt{\frac{b^2+ 4\Delta^2}{b^2 +  4\Delta^2 (\delta h_{z}/\delta h_{xy})^2}},
\end{equation}
with $T_2^* = \sqrt{2} \hbar/\sigma$ the typical inhomogeneous dephasing time associated with nuclear spins. Equation~(\ref{P_detuning}) implies a crossover between the ESR power-law decay at weak drive to the gaussian decay of the strong-drive regime. 

To exemplify this behavior, we first consider the resonant condition ($\Delta =0$) when, besides $\delta R/\delta x$, the form of the decay is determined by $T_2^*$ and the coefficient $\eta$. An example of numerical results  for $\overline{P_\downarrow(t )}$ is shown in Fig.~\ref{fig:1}(b), assuming $\eta=0.05$. We confirm that the ESR power-law decay \cite{Koppens2006,Koppens2007} is recovered when $\delta R/\delta x \to 0$ but significant deviations from this known dependence appear at larger strength of the drive. In the large-drive limit, we find a gaussian decay with no universal $\pi/4$ phase shift, as Eq.~(\ref{P_detuning}) becomes an excellent approximation. This crossover to the strong-drive regime occurs when the effect of $h_x$  in Eq.~(\ref{Rabi_no_noise}) becomes dominant over $h_z$, i.e., $b h_x \gg h_z^2$. We estimate the typical values of $h_{x},h_z$ using the limit of $\delta h_{xy},\delta h_z$ at small $\delta R/\delta x$ (which is justified if $\eta \ll 1$), i.e., $h_{x}\sim \frac12 \sigma \delta R/\delta x $ and $h_z \sim \sigma$. This yields the condition:
\begin{equation}\label{gaussian_condition}
\frac{\delta R}{\delta x} \gtrsim \sqrt{2 \eta},
\end{equation}
in good agreement with the numerical results of Fig.~\ref{fig:1}(b).

Of special interest is the decay timescale:
\begin{equation}\label{TR0}
T_R(\Delta=0)= \frac{T_2^* }{\delta h_{xy}},
\end{equation}
which follows simply from Eq.~(\ref{TR0general}) and is plotted in Fig.~\ref{fig:1}(c). The increasing strength of $\delta h_{xy}$ with the drive leads to a significantly faster decay, as also seen in the time domain results of Fig. \ref{fig:1}(b). However, it is important to note that the fidelity of the $\pi$ rotation grows monotonically with $\delta R$, as shown in Fig.~\ref{fig:1}(d). For a quantitative analysis, we set $t=2 \pi \hbar/b $ in Eq.~(\ref{Rabi_no_noise}) and perform an expansion up to second order in $\vec{h}$. After statistical averaging, we have:
\begin{equation}\label{infidelity}
1-\max[\overline{P_\downarrow (t)} ] \simeq  \frac{\eta^2  ( 4 \delta h_z^2 + \pi^2 \delta h_{xy}^2)}{(\delta R / \delta x)^2} < \frac{4.1 \eta^2 }{(\delta R / \delta x)^2}.
\end{equation}
Equation~(\ref{infidelity}) includes a contribution proportional to $\delta h_{xy}^2$ but the additional dephasing from transverse fluctuations is more than compensated by the decrease of $\delta h_z^2$ and the faster Rabi frequency. Thus, Eq.~(\ref{infidelity}) shows that it is always advantageous to apply a stronger drive and the effect of the hyperfine interaction on $\pi$-rotation error can be reduced below any desired threshold with a sufficiently large $\delta R /\delta x$ \cite{Yoneda2014}.

\begin{figure}
\begin{center}
\includegraphics[height=0.19\textwidth]{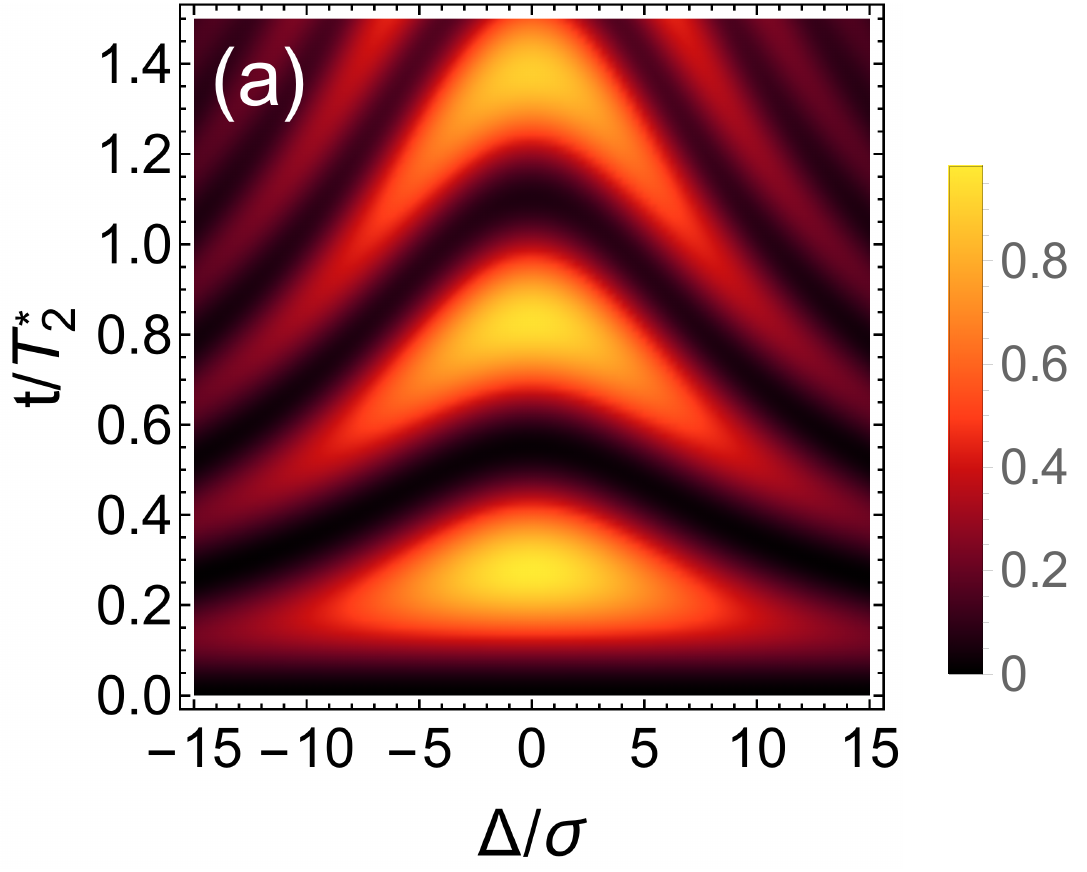}
\includegraphics[height=0.19\textwidth]{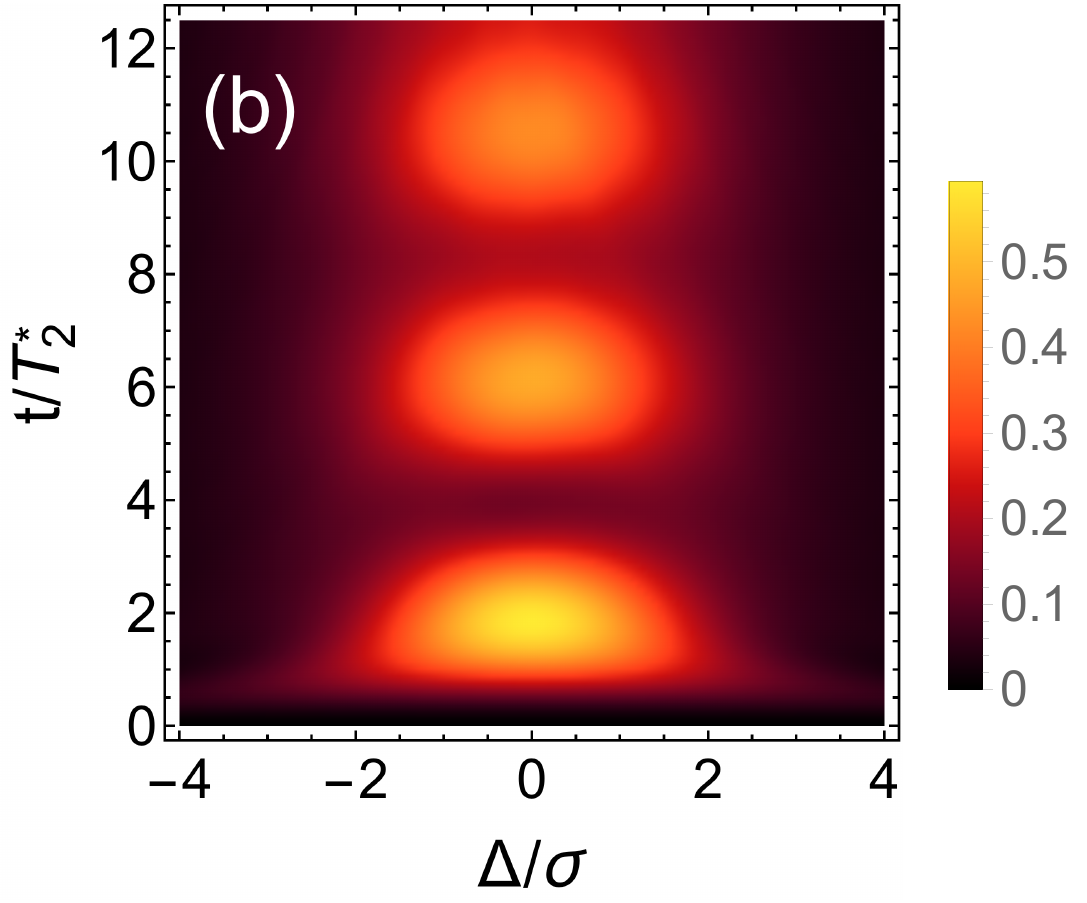}
\includegraphics[height=0.22\textwidth]{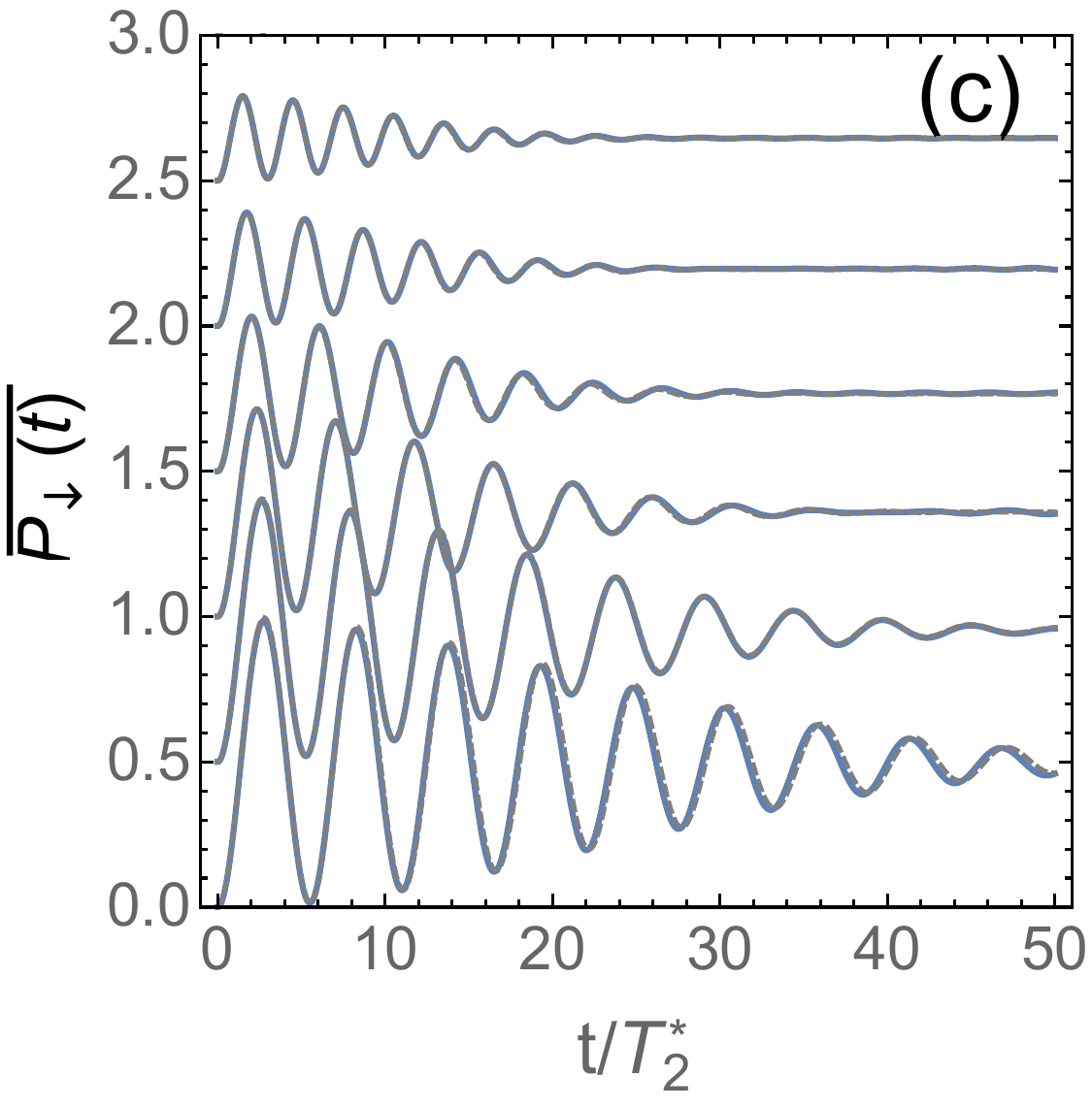}
\includegraphics[height=0.22\textwidth]{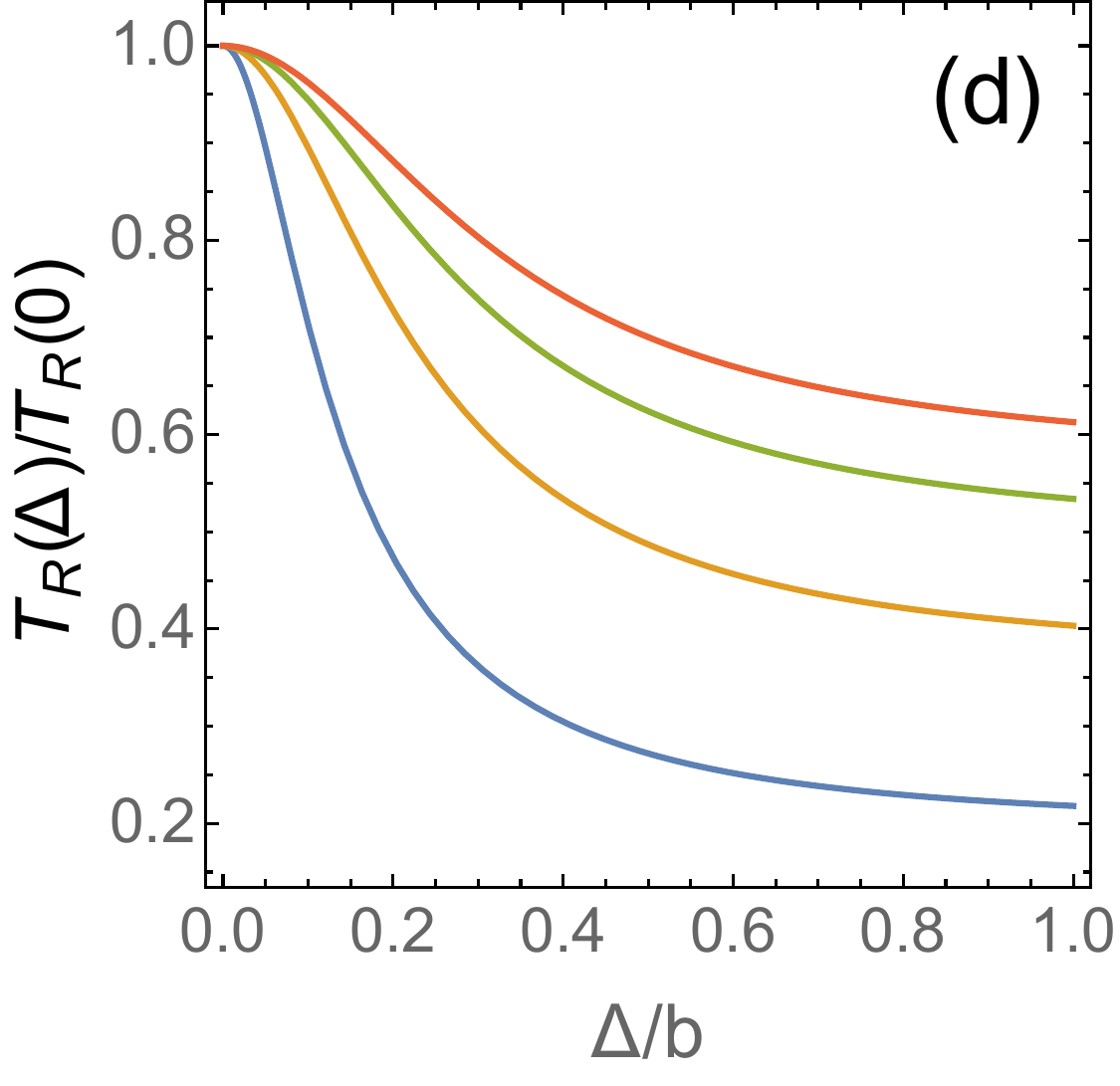}
\caption{Rabi oscillations at finite detuning. (a) and (b) are for $\delta R/\delta x= 0.8$ and 0.1, respectively. In both panels, $\eta =0.05$. (c) shows the line cuts of (a) at $\Delta/\sigma = 0,2.5,5, \ldots 12.5$ (bottom to top). For clarity, the curves are shifted vertically. The numerical results are virtually indistinguishable from Eq.~(\ref{P_detuning}), plotted as grey dashed curves. In panel (d) we plot Eq.~(\ref{TR0general}) as function of detuning for several values of $\delta R/\delta x = 0.4,0.8,1.2,1.6$ (bottom to top).
\label{fig:2}}
\end{center}
\end{figure}

\paragraph{Decay at finite detuning.} Considering a finite detuning yields further insight on the role of longitudinal/transverse nuclear fluctuations. The effect of $\Delta$ is illustrated in Fig.~\ref{fig:2}, where panel (c) confirms that Eq.~(\ref{P_detuning}) provides an excellent approximation in the strong drive regime. A first consequence is that the Rabi oscillations approach the ``chevron'' pattern of Fig.~\ref{fig:2}(a). Furthermore the decay time gets reduced at finite $\Delta$, which is illustrated in Figs.~\ref{fig:2}(c) and (d). 

The dependence of $T_R$ on $\Delta$ has a simple physical explanation, as it can be traced to the difference in strength between transverse and longitudinal nuclear fluctuations shown in Fig.~\ref{fig:1}(a). Since Eq.~(\ref{Hsimple}) implies that a finite detuning corresponds to a field along $z$ in the rotating frame, the relevant component of the nuclear fluctuations (i.e., along the total driving field) becomes a weighted average of $h_x$ and $h_z$. Since $\delta h_z > \delta h_{xy}$, the nuclear fluctuations gets enhanced by a finite detuning. As shown in Fig.~\ref{fig:2}(d), the dependence of $T_R$ on $\Delta$ is particularly pronounced at smaller values of $\delta R/\delta x$. This is natural, as the ratio $\delta h_z/\delta h_{xy}$ is large in this regime, see Fig.~\ref{fig:1}(a), while the nuclear fluctuations become more isotropic at larger $\delta R/\delta x$. Thus, studying the dependence of $T_R$ on $\Delta$ allows one to explore how the relative strength of longitudinal/transverse nuclear fluctuations evolves with $\delta R/\delta x$. 

\paragraph{Stationary limit.} We conclude our analysis of the Rabi oscillations by commenting briefly on the stationary limit $\overline{P_\downarrow(t\to \infty)}$, which is a useful quantity to estimate $\sigma$.  Common methods rely either on the drive dependence at resonance (as done in the ESR experiment of Ref.~\cite{Koppens2007}) or the linewidth (considering finite detunings).  We find that, even if the Rabi oscillations are sensitively modified by transverse nuclear fluctuations, the effect on $\overline{P_\downarrow(\infty)}$ is negligible in the current experimental regime $\eta \ll 1$.  A significant difference between ESR and EDSR only appears when $\eta  \sim 1$ (for more detais, see \cite{supplemental}). Therefore, methods to extract $\sigma$ from  $\overline{P_\downarrow(\infty)}$ are still valid for large-amplitude EDSR. 

\begin{figure}
\begin{center}
\includegraphics[height=0.225\textwidth]{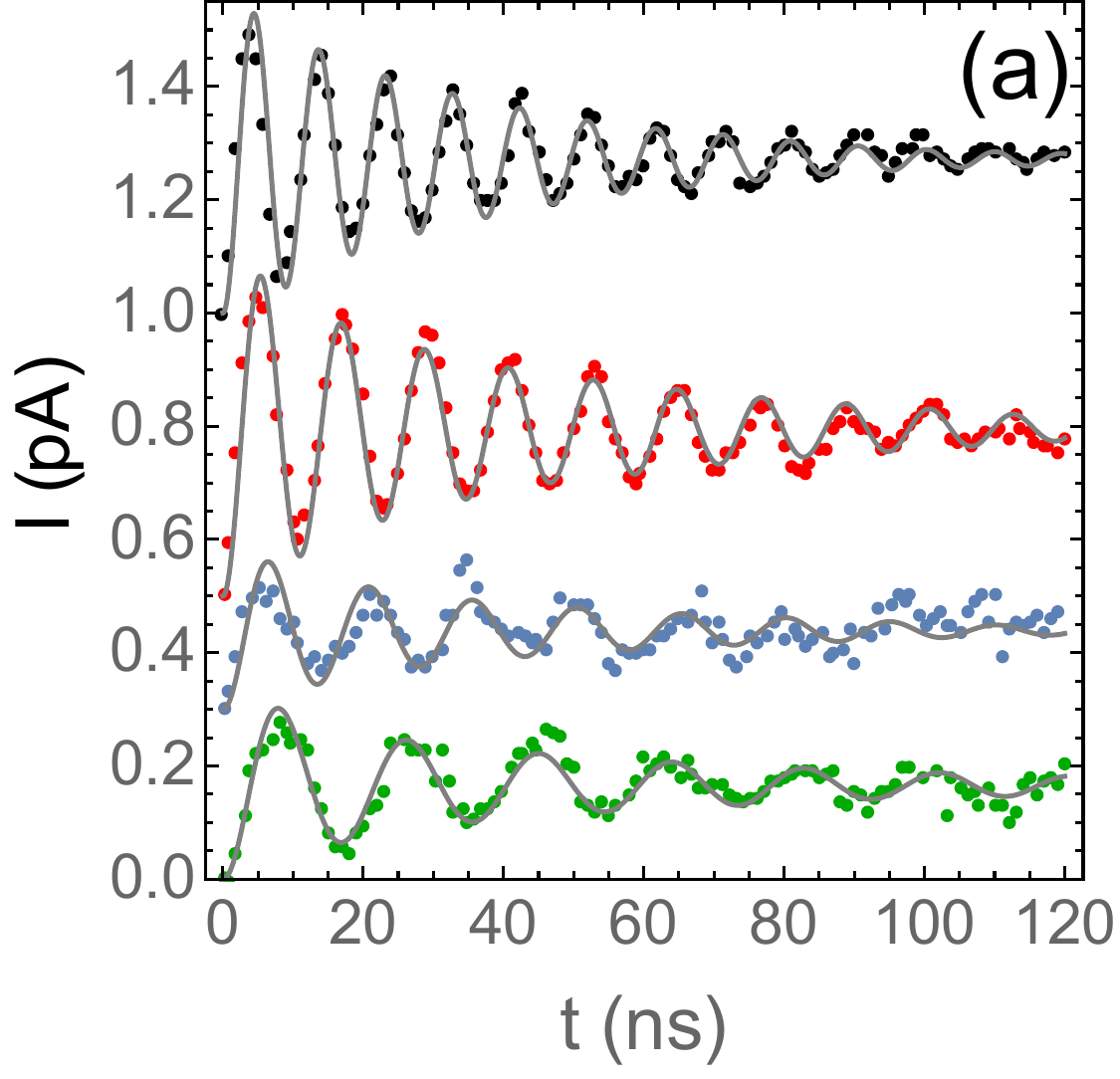}
\includegraphics[height=0.225\textwidth]{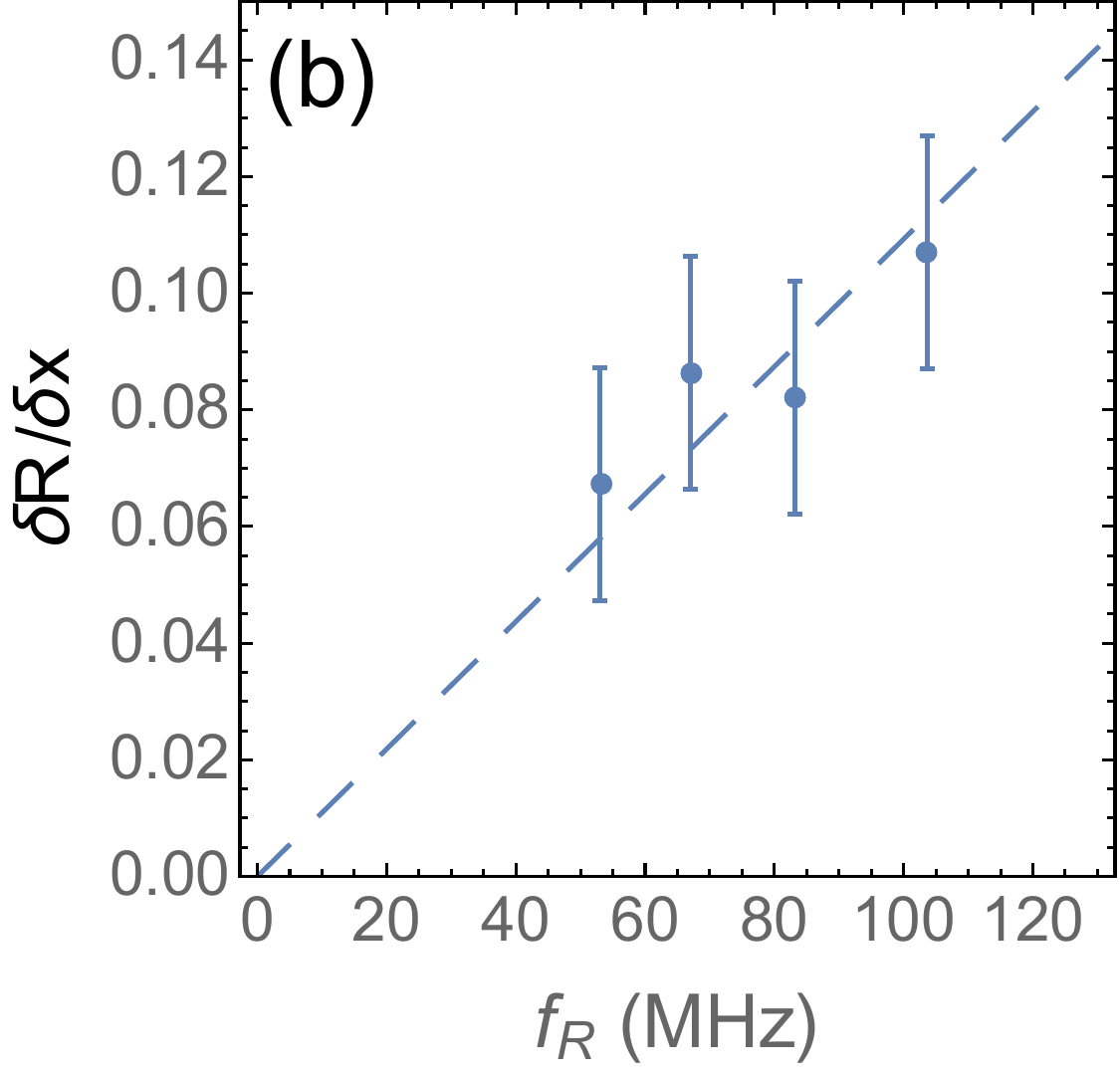}
\caption{(a): Fit of our theory to Rabi oscillations of Ref.~\cite{vdBerg2013}. For each curve, the fit parameters are $\delta R/\delta x$, a conversion factor to current, and $b = 4 \pi \hbar f_R$, while $\sigma = 0.22 \pm 0.03 ~\mu{\rm eV}$ is from the experiment \cite{vdBerg2013}. (b): $\delta R/\delta x$ and $f_R$ obtained from the fits in (a). The error bars are from the uncertainty on $\sigma$. The dashed line is a fit to $\delta R/\delta x = C f_R$.
\label{fig:3}}
\end{center}
\end{figure}

\paragraph{Comparison to experiments.} We now discuss the application of our theory to available experimental data. In Fig.~\ref{fig:3} we show an analysis of the data shown in Fig.~2(c) of Ref.~\cite{vdBerg2013}, obtained from InSb nanowire dots with strong spin-orbit interaction. As seen, our theory is able to reproduce well the Rabi oscillations and the fit yields values $\delta R/\delta x$ and $b$ consistent with $b \propto \delta R/\delta x$. For EDSR driven by the spin-orbit coupling we have:
\begin{equation}
\eta = \frac{l_{SO}}{2\delta x} \, \frac{\sigma}{\epsilon_z \sin\theta},
\end{equation} 
where $l_{SO}$ is the spin-orbit length and $\theta$ the angle between the spin-orbit field and $\vec B$. Using $l_{SO} =200 - 300 $ nm, $\delta x \simeq 10 - 20$ nm \cite{Nadj-Perge2012},  $g=41$, and $B=31.4$~mT \cite{vdBerg2013} gives $\eta \simeq (0.015 - 0.04)/\sin\theta$. Figure~\ref{fig:3}(b) implies $\eta \sim 0.03$, which is consistent with this estimate. 

While Fig.~\ref{fig:3}(b) has $\delta R / \delta x \lesssim 0.1$, we estimate that larger values of $\delta R / \delta x$ were achieved in a recent experiment on GaAs quantum dots \cite{Yoneda2014}. There, the drive is based on a micromagnet for which numerical simulations give $b/\delta R \sim|g|\mu_B \times$(1 mT/nm) \cite{Yoneda2014}. Using the largest achieved Rabi frequency $f_{\rm max} \sim 120$~MHz, $|g| \simeq 0.4$, and $\delta x \sim 35-60$~nm (corresponding to orbital energies $\sim 0.3-1$~meV), we obtain $\delta R_{\rm max}/\delta x \sim 0.7 - 1.2$ which is relatively close to the condition at which $h_x$ fluctuations are most effective. On the other hand, the regime of motional narrowing $\delta R / \delta x \gtrsim 1.8$ \cite{comment_motional_narrowing} does not appear to be within reach of current experiments. For GaAs quantum dots $\sigma \sim |g|\mu_B \times (1-4~{\rm mT})$ \cite{Petta2005,Koppens2006,Koppens2007,Laird2007}, which allows us to estimate a typical range $\eta \sim 0.02 - 0.1$.

Several findings of Ref.~\cite{Yoneda2014} are in good agreement with our discussion, including the transition to a ``chevron'' pattern in the strong drive regime \cite{Yoneda2014}. As a result, our panels (a) and (b) of Fig.~\ref{fig:2} are remarkably similar to the corresponding panels of Ref.~\cite{Yoneda2014}.  Furthermore, a crossover from power-law decay (for $b \lesssim |g|\mu_B\times$ 5~mT) to gaussian decay (for $b \gtrsim |g|\mu_B \times$ 15~mT) was observed. The strength of the drive for such crossover is compatible with Eq.~(\ref{gaussian_condition}), which can be rewritten as $b \gtrsim \sqrt{2\sigma \delta x (b/\delta R)} \sim |g|\mu_B \times (8 - 20 ~{\rm mT})$, using the above estimates of $\sigma, \delta x$, and $b/\delta R$. We also show in Fig.~\ref{fig:1}(c) that Eq.~(\ref{TR0}) is able to reproduce the dependence of $T_R$ on the drive strength, with reasonable fit parameters for GaAs quantum dots ($T_2^*\simeq 9$ ns, $\eta \simeq 0.09$). 

\paragraph{Conclusion.} In conclusion, we have characterized the dephasing induced by the hypefine interaction in large-amplitude EDSR, and we showed that transverse fluctuations of the Overhauser field are likely to play an important role in this regime, recently achieved experimentally. It should be mentioned that also other dephasing sources were suggested for EDSR, such as paramagnetic impurities, charge noise, and photon-assisted tunneling \cite{NadjPerge2010,vdBerg2013,Yoneda2014}. In the absence of clear evidence (or specific predictions) about these alternative mechanisms, our theory offers further means to test if nuclear spins are indeed the dominant effect, e.g., through a detailed analysis of $T_R$ as function of both drive and detuning. In fact, it is unlikely that other types of dephasing would induce the same type of sensitive dependence on $\Delta$ discussed in relation to Eq.~(\ref{TR0general}) and Fig.~\ref{fig:2}.

From a more general point of view, our study is helpful to assess the limitations to spin manipulation due to the quantum-dot motion and the nuclear-spin bath. These two aspects are unavoidable for EDSR based on III-V semiconductors, in contrast to ESR or spin manipulation based on group-IV materials \cite{Kawakami2014,Veldhorst2014}. Nuclear fluctuations, on the other hand, do not represent a fundamental obstacle to EDSR, since high-fidelity gates can be achieved at sufficiently large amplitude.

We thank W. A. Coish, T. Otsuka, P. Stano, and J. Yoneda for useful discussions. S. C. acknowledges funding from NSFC (grant No. 11574025). D. L. acknowledges support from the Swiss NSF and NCCR QSIT.

\bibliography{bibfile}

\begin{thebibliography}{25}%
\makeatletter
\providecommand \@ifxundefined [1]{%
 \@ifx{#1\undefined}
}%
\providecommand \@ifnum [1]{%
 \ifnum #1\expandafter \@firstoftwo
 \else \expandafter \@secondoftwo
 \fi
}%
\providecommand \@ifx [1]{%
 \ifx #1\expandafter \@firstoftwo
 \else \expandafter \@secondoftwo
 \fi
}%
\providecommand \natexlab [1]{#1}%
\providecommand \enquote  [1]{``#1''}%
\providecommand \bibnamefont  [1]{#1}%
\providecommand \bibfnamefont [1]{#1}%
\providecommand \citenamefont [1]{#1}%
\providecommand \href@noop [0]{\@secondoftwo}%
\providecommand \href [0]{\begingroup \@sanitize@url \@href}%
\providecommand \@href[1]{\@@startlink{#1}\@@href}%
\providecommand \@@href[1]{\endgroup#1\@@endlink}%
\providecommand \@sanitize@url [0]{\catcode `\\12\catcode `\$12\catcode
  `\&12\catcode `\#12\catcode `\^12\catcode `\_12\catcode `\%12\relax}%
\providecommand \@@startlink[1]{}%
\providecommand \@@endlink[0]{}%
\providecommand \url  [0]{\begingroup\@sanitize@url \@url }%
\providecommand \@url [1]{\endgroup\@href {#1}{\urlprefix }}%
\providecommand \urlprefix  [0]{URL }%
\providecommand \Eprint [0]{\href }%
\providecommand \doibase [0]{http://dx.doi.org/}%
\providecommand \selectlanguage [0]{\@gobble}%
\providecommand \bibinfo  [0]{\@secondoftwo}%
\providecommand \bibfield  [0]{\@secondoftwo}%
\providecommand \translation [1]{[#1]}%
\providecommand \BibitemOpen [0]{}%
\providecommand \bibitemStop [0]{}%
\providecommand \bibitemNoStop [0]{.\EOS\space}%
\providecommand \EOS [0]{\spacefactor3000\relax}%
\providecommand \BibitemShut  [1]{\csname bibitem#1\endcsname}%
\let\auto@bib@innerbib\@empty
\bibitem [{\citenamefont {Hanson}\ \emph {et~al.}(2007)\citenamefont {Hanson},
  \citenamefont {Petta}, \citenamefont {Tarucha},\ and\ \citenamefont
  {Vandersypen}}]{Hanson2007}%
  \BibitemOpen
  \bibfield  {author} {\bibinfo {author} {\bibfnamefont {R.}~\bibnamefont
  {Hanson}}, \bibinfo {author} {\bibfnamefont {J.~R.}\ \bibnamefont {Petta}},
  \bibinfo {author} {\bibfnamefont {S.}~\bibnamefont {Tarucha}}, \ and\
  \bibinfo {author} {\bibfnamefont {L.~M.~K.}\ \bibnamefont {Vandersypen}},\
  }\href@noop {} {\bibfield  {journal} {\bibinfo  {journal} {Rev. Mod. Phys.}\
  }\textbf {\bibinfo {volume} {79}},\ \bibinfo {pages} {1217} (\bibinfo {year}
  {2007})}\BibitemShut {NoStop}%
\bibitem [{\citenamefont {Awschalom}\ \emph {et~al.}(2013)\citenamefont
  {Awschalom}, \citenamefont {Bassett}, \citenamefont {Dzurak}, \citenamefont
  {Hu},\ and\ \citenamefont {Petta}}]{Awschalom2013}%
  \BibitemOpen
  \bibfield  {author} {\bibinfo {author} {\bibfnamefont {D.~D.}\ \bibnamefont
  {Awschalom}}, \bibinfo {author} {\bibfnamefont {L.~C.}\ \bibnamefont
  {Bassett}}, \bibinfo {author} {\bibfnamefont {A.~S.}\ \bibnamefont {Dzurak}},
  \bibinfo {author} {\bibfnamefont {E.~L.}\ \bibnamefont {Hu}}, \ and\ \bibinfo
  {author} {\bibfnamefont {J.~R.}\ \bibnamefont {Petta}},\ }\href@noop {}
  {\bibfield  {journal} {\bibinfo  {journal} {Science}\ }\textbf {\bibinfo
  {volume} {339}},\ \bibinfo {pages} {1174} (\bibinfo {year}
  {2013})}\BibitemShut {NoStop}%
\bibitem [{\citenamefont {Koppens}\ \emph {et~al.}(2006)\citenamefont
  {Koppens}, \citenamefont {Buizert}, \citenamefont {Tielrooij}, \citenamefont
  {Vink}, \citenamefont {Nowack}, \citenamefont {Meunier}, \citenamefont
  {Kouwenhoven},\ and\ \citenamefont {Vandersypen}}]{Koppens2006}%
  \BibitemOpen
  \bibfield  {author} {\bibinfo {author} {\bibfnamefont {F.~H.~L.}\
  \bibnamefont {Koppens}}, \bibinfo {author} {\bibfnamefont {C.}~\bibnamefont
  {Buizert}}, \bibinfo {author} {\bibfnamefont {K.~J.}\ \bibnamefont
  {Tielrooij}}, \bibinfo {author} {\bibfnamefont {I.~T.}\ \bibnamefont {Vink}},
  \bibinfo {author} {\bibfnamefont {K.~C.}\ \bibnamefont {Nowack}}, \bibinfo
  {author} {\bibfnamefont {T.}~\bibnamefont {Meunier}}, \bibinfo {author}
  {\bibfnamefont {L.~P.}\ \bibnamefont {Kouwenhoven}}, \ and\ \bibinfo {author}
  {\bibfnamefont {L.~M.}\ \bibnamefont {Vandersypen}},\ }\href@noop {}
  {\bibfield  {journal} {\bibinfo  {journal} {Nature (London)}\ }\textbf
  {\bibinfo {volume} {442}},\ \bibinfo {pages} {766} (\bibinfo {year}
  {2006})}\BibitemShut {NoStop}%
\bibitem [{\citenamefont {Loss}\ and\ \citenamefont
  {DiVincenzo}(1998)}]{Loss1998}%
  \BibitemOpen
  \bibfield  {author} {\bibinfo {author} {\bibfnamefont {D.}~\bibnamefont
  {Loss}}\ and\ \bibinfo {author} {\bibfnamefont {D.~P.}\ \bibnamefont
  {DiVincenzo}},\ }\href@noop {} {\bibfield  {journal} {\bibinfo  {journal}
  {Phys. Rev. A}\ }\textbf {\bibinfo {volume} {57}},\ \bibinfo {pages} {120}
  (\bibinfo {year} {1998})}\BibitemShut {NoStop}%
\bibitem [{\citenamefont {Golovach}\ \emph {et~al.}(2006)\citenamefont
  {Golovach}, \citenamefont {Borhani},\ and\ \citenamefont
  {Loss}}]{Golovach2006}%
  \BibitemOpen
  \bibfield  {author} {\bibinfo {author} {\bibfnamefont {V.~N.}\ \bibnamefont
  {Golovach}}, \bibinfo {author} {\bibfnamefont {M.}~\bibnamefont {Borhani}}, \
  and\ \bibinfo {author} {\bibfnamefont {D.}~\bibnamefont {Loss}},\ }\href@noop
  {} {\bibfield  {journal} {\bibinfo  {journal} {Phys. Rev. B}\ }\textbf
  {\bibinfo {volume} {74}},\ \bibinfo {pages} {165319} (\bibinfo {year}
  {2006})}\BibitemShut {NoStop}%
\bibitem [{\citenamefont {Nowack}\ \emph {et~al.}(2007)\citenamefont {Nowack},
  \citenamefont {Koppens}, \citenamefont {Nazarov},\ and\ \citenamefont
  {Vandersypen}}]{Nowack2007}%
  \BibitemOpen
  \bibfield  {author} {\bibinfo {author} {\bibfnamefont {K.~C.}\ \bibnamefont
  {Nowack}}, \bibinfo {author} {\bibfnamefont {F.~H.~L.}\ \bibnamefont
  {Koppens}}, \bibinfo {author} {\bibfnamefont {Y.~V.}\ \bibnamefont
  {Nazarov}}, \ and\ \bibinfo {author} {\bibfnamefont {L.~M.~K.}\ \bibnamefont
  {Vandersypen}},\ }\href {\doibase 10.1126/science.1148092} {\bibfield
  {journal} {\bibinfo  {journal} {Science}\ }\textbf {\bibinfo {volume}
  {318}},\ \bibinfo {pages} {1430} (\bibinfo {year} {2007})}\BibitemShut
  {NoStop}%
\bibitem [{\citenamefont {Tokura}\ \emph {et~al.}(2006)\citenamefont {Tokura},
  \citenamefont {van~der Wiel}, \citenamefont {Obata},\ and\ \citenamefont
  {Tarucha}}]{Tokura2006}%
  \BibitemOpen
  \bibfield  {author} {\bibinfo {author} {\bibfnamefont {Y.}~\bibnamefont
  {Tokura}}, \bibinfo {author} {\bibfnamefont {W.~G.}\ \bibnamefont {van~der
  Wiel}}, \bibinfo {author} {\bibfnamefont {T.}~\bibnamefont {Obata}}, \ and\
  \bibinfo {author} {\bibfnamefont {S.}~\bibnamefont {Tarucha}},\ }\href
  {\doibase 10.1103/PhysRevLett.96.047202} {\bibfield  {journal} {\bibinfo
  {journal} {Phys. Rev. Lett.}\ }\textbf {\bibinfo {volume} {96}},\ \bibinfo
  {pages} {047202} (\bibinfo {year} {2006})}\BibitemShut {NoStop}%
\bibitem [{\citenamefont {Pioro-Ladri\`ere}\ \emph {et~al.}(2008)\citenamefont
  {Pioro-Ladri\`ere}, \citenamefont {Obata}, \citenamefont {Tokura},
  \citenamefont {Shin}, \citenamefont {Kubo}, \citenamefont {Yoshida},
  \citenamefont {Taniyama},\ and\ \citenamefont {Tarucha}}]{Pioro2008}%
  \BibitemOpen
  \bibfield  {author} {\bibinfo {author} {\bibfnamefont {M.}~\bibnamefont
  {Pioro-Ladri\`ere}}, \bibinfo {author} {\bibfnamefont {T.}~\bibnamefont
  {Obata}}, \bibinfo {author} {\bibfnamefont {Y.}~\bibnamefont {Tokura}},
  \bibinfo {author} {\bibfnamefont {Y.-S.}\ \bibnamefont {Shin}}, \bibinfo
  {author} {\bibfnamefont {T.}~\bibnamefont {Kubo}}, \bibinfo {author}
  {\bibfnamefont {K.}~\bibnamefont {Yoshida}}, \bibinfo {author} {\bibfnamefont
  {T.}~\bibnamefont {Taniyama}}, \ and\ \bibinfo {author} {\bibfnamefont
  {S.}~\bibnamefont {Tarucha}},\ }\href@noop {} {\bibfield  {journal} {\bibinfo
   {journal} {Nature Phys.}\ }\textbf {\bibinfo {volume} {4}},\ \bibinfo
  {pages} {776} (\bibinfo {year} {2008})}\BibitemShut {NoStop}%
\bibitem [{\citenamefont {van~den Berg}\ \emph {et~al.}(2013)\citenamefont
  {van~den Berg}, \citenamefont {Nadj-Perge}, \citenamefont {Pribiag},
  \citenamefont {Plissard}, \citenamefont {Bakkers}, \citenamefont {Frolov},\
  and\ \citenamefont {Kouwenhoven}}]{vdBerg2013}%
  \BibitemOpen
  \bibfield  {author} {\bibinfo {author} {\bibfnamefont {J.~W.~G.}\
  \bibnamefont {van~den Berg}}, \bibinfo {author} {\bibfnamefont
  {S.}~\bibnamefont {Nadj-Perge}}, \bibinfo {author} {\bibfnamefont {V.~S.}\
  \bibnamefont {Pribiag}}, \bibinfo {author} {\bibfnamefont {S.~R.}\
  \bibnamefont {Plissard}}, \bibinfo {author} {\bibfnamefont {E.~P. A.~M.}\
  \bibnamefont {Bakkers}}, \bibinfo {author} {\bibfnamefont {S.~M.}\
  \bibnamefont {Frolov}}, \ and\ \bibinfo {author} {\bibfnamefont {L.~P.}\
  \bibnamefont {Kouwenhoven}},\ }\href@noop {} {\bibfield  {journal} {\bibinfo
  {journal} {Phys. Rev. Lett.}\ }\textbf {\bibinfo {volume} {110}},\ \bibinfo
  {pages} {066806} (\bibinfo {year} {2013})}\BibitemShut {NoStop}%
\bibitem [{\citenamefont {Yoneda}\ \emph {et~al.}(2014)\citenamefont {Yoneda},
  \citenamefont {Otsuka}, \citenamefont {Nakajima}, \citenamefont {Takakura},
  \citenamefont {Obata}, \citenamefont {Pioro-Ladri\`ere}, \citenamefont {Lu},
  \citenamefont {Palmstr{\o}m}, \citenamefont {Gossard},\ and\ \citenamefont
  {Tarucha}}]{Yoneda2014}%
  \BibitemOpen
  \bibfield  {author} {\bibinfo {author} {\bibfnamefont {J.}~\bibnamefont
  {Yoneda}}, \bibinfo {author} {\bibfnamefont {T.}~\bibnamefont {Otsuka}},
  \bibinfo {author} {\bibfnamefont {T.}~\bibnamefont {Nakajima}}, \bibinfo
  {author} {\bibfnamefont {T.}~\bibnamefont {Takakura}}, \bibinfo {author}
  {\bibfnamefont {T.}~\bibnamefont {Obata}}, \bibinfo {author} {\bibfnamefont
  {M.}~\bibnamefont {Pioro-Ladri\`ere}}, \bibinfo {author} {\bibfnamefont
  {H.}~\bibnamefont {Lu}}, \bibinfo {author} {\bibfnamefont {C.~J.}\
  \bibnamefont {Palmstr{\o}m}}, \bibinfo {author} {\bibfnamefont {A.~C.}\
  \bibnamefont {Gossard}}, \ and\ \bibinfo {author} {\bibfnamefont
  {S.}~\bibnamefont {Tarucha}},\ }\href@noop {} {\bibfield  {journal} {\bibinfo
   {journal} {Phys. Rev. Lett.}\ }\textbf {\bibinfo {volume} {113}},\ \bibinfo
  {pages} {267601} (\bibinfo {year} {2014})}\BibitemShut {NoStop}%
\bibitem [{\citenamefont {Koppens}\ \emph {et~al.}(2007)\citenamefont
  {Koppens}, \citenamefont {Klauser}, \citenamefont {Coish}, \citenamefont
  {Nowack}, \citenamefont {Kouwenhoven}, \citenamefont {Loss},\ and\
  \citenamefont {Vandersypen}}]{Koppens2007}%
  \BibitemOpen
  \bibfield  {author} {\bibinfo {author} {\bibfnamefont {F.~H.~L.}\
  \bibnamefont {Koppens}}, \bibinfo {author} {\bibfnamefont {D.}~\bibnamefont
  {Klauser}}, \bibinfo {author} {\bibfnamefont {W.~A.}\ \bibnamefont {Coish}},
  \bibinfo {author} {\bibfnamefont {K.~C.}\ \bibnamefont {Nowack}}, \bibinfo
  {author} {\bibfnamefont {L.~P.}\ \bibnamefont {Kouwenhoven}}, \bibinfo
  {author} {\bibfnamefont {D.}~\bibnamefont {Loss}}, \ and\ \bibinfo {author}
  {\bibfnamefont {L.~M.~K.}\ \bibnamefont {Vandersypen}},\ }\href {\doibase
  10.1103/PhysRevLett.99.106803} {\bibfield  {journal} {\bibinfo  {journal}
  {Phys. Rev. Lett.}\ }\textbf {\bibinfo {volume} {99}},\ \bibinfo {pages}
  {106803} (\bibinfo {year} {2007})}\BibitemShut {NoStop}%
\bibitem [{\citenamefont {Nadj-Perge}\ \emph {et~al.}(2010)\citenamefont
  {Nadj-Perge}, \citenamefont {Frolov}, \citenamefont {Bakkers},\ and\
  \citenamefont {Kouwenhoven}}]{NadjPerge2010}%
  \BibitemOpen
  \bibfield  {author} {\bibinfo {author} {\bibfnamefont {S.}~\bibnamefont
  {Nadj-Perge}}, \bibinfo {author} {\bibfnamefont {S.~M.}\ \bibnamefont
  {Frolov}}, \bibinfo {author} {\bibfnamefont {E.~P. A.~M.}\ \bibnamefont
  {Bakkers}}, \ and\ \bibinfo {author} {\bibfnamefont {L.~P.}\ \bibnamefont
  {Kouwenhoven}},\ }\href@noop {} {\bibfield  {journal} {\bibinfo  {journal}
  {Nature}\ }\textbf {\bibinfo {volume} {468}},\ \bibinfo {pages} {1084}
  (\bibinfo {year} {2010})}\BibitemShut {NoStop}%
\bibitem [{\citenamefont {Laird}\ \emph {et~al.}(2007)\citenamefont {Laird},
  \citenamefont {Barthel}, \citenamefont {Rashba}, \citenamefont {Marcus},
  \citenamefont {Hanson},\ and\ \citenamefont {Gossard}}]{Laird2007}%
  \BibitemOpen
  \bibfield  {author} {\bibinfo {author} {\bibfnamefont {E.~A.}\ \bibnamefont
  {Laird}}, \bibinfo {author} {\bibfnamefont {C.}~\bibnamefont {Barthel}},
  \bibinfo {author} {\bibfnamefont {E.~I.}\ \bibnamefont {Rashba}}, \bibinfo
  {author} {\bibfnamefont {C.~M.}\ \bibnamefont {Marcus}}, \bibinfo {author}
  {\bibfnamefont {M.~P.}\ \bibnamefont {Hanson}}, \ and\ \bibinfo {author}
  {\bibfnamefont {A.~C.}\ \bibnamefont {Gossard}},\ }\href@noop {} {\bibfield
  {journal} {\bibinfo  {journal} {Phys. Rev. Lett.}\ }\textbf {\bibinfo
  {volume} {99}},\ \bibinfo {pages} {246601} (\bibinfo {year}
  {2007})}\BibitemShut {NoStop}%
\bibitem [{\citenamefont {Rashba}(2008)}]{Rashba2008}%
  \BibitemOpen
  \bibfield  {author} {\bibinfo {author} {\bibfnamefont {E.~I.}\ \bibnamefont
  {Rashba}},\ }\href@noop {} {\bibfield  {journal} {\bibinfo  {journal} {Phys.
  Rev. B}\ }\textbf {\bibinfo {volume} {78}},\ \bibinfo {pages} {195302}
  (\bibinfo {year} {2008})}\BibitemShut {NoStop}%
\bibitem [{\citenamefont {Sz\'echenyi}\ and\ \citenamefont
  {P\'alyi}(2014)}]{Palyi2014}%
  \BibitemOpen
  \bibfield  {author} {\bibinfo {author} {\bibfnamefont {G.}~\bibnamefont
  {Sz\'echenyi}}\ and\ \bibinfo {author} {\bibfnamefont {A.}~\bibnamefont
  {P\'alyi}},\ }\href@noop {} {\bibfield  {journal} {\bibinfo  {journal} {Phys.
  Rev. B}\ }\textbf {\bibinfo {volume} {89}},\ \bibinfo {pages} {115409}
  (\bibinfo {year} {2014})}\BibitemShut {NoStop}%
\bibitem [{\citenamefont {Echeverr\'{\i}a-Arrondo}\ and\ \citenamefont
  {Sherman}(2013)}]{Arrondo2013}%
  \BibitemOpen
  \bibfield  {author} {\bibinfo {author} {\bibfnamefont {C.}~\bibnamefont
  {Echeverr\'{\i}a-Arrondo}}\ and\ \bibinfo {author} {\bibfnamefont {E.~Y.}\
  \bibnamefont {Sherman}},\ }\href@noop {} {\bibfield  {journal} {\bibinfo
  {journal} {Phys. Rev. B}\ }\textbf {\bibinfo {volume} {87}},\ \bibinfo
  {pages} {081410} (\bibinfo {year} {2013})}\BibitemShut {NoStop}%
\bibitem [{\citenamefont {Jing}\ \emph {et~al.}(2014)\citenamefont {Jing},
  \citenamefont {Huang},\ and\ \citenamefont {Hu}}]{Jing2014}%
  \BibitemOpen
  \bibfield  {author} {\bibinfo {author} {\bibfnamefont {J.}~\bibnamefont
  {Jing}}, \bibinfo {author} {\bibfnamefont {P.}~\bibnamefont {Huang}}, \ and\
  \bibinfo {author} {\bibfnamefont {X.}~\bibnamefont {Hu}},\ }\href@noop {}
  {\bibfield  {journal} {\bibinfo  {journal} {Phys. Rev. A}\ }\textbf {\bibinfo
  {volume} {90}},\ \bibinfo {pages} {022118} (\bibinfo {year}
  {2014})}\BibitemShut {NoStop}%
\bibitem [{rem()}]{remark_by_bz}%
  \BibitemOpen
  \href@noop {} {}\bibinfo {note} {For an infinite temperature nuclear bath, we
  can always choose the drive along $x$ in spin space due to the rotational
  invariance of the hyperfine coupling in the $x-y$ plane, see
  Eq.~(\ref{Hrot}). The strength of the drive is $b_\perp =\sqrt{b_x^2
  +b_y^2}=b$ if $b_z=0$. If $b_z \neq 0$, one should substitute $b \to b_\perp$
  in Eq.~(\ref{Hsimple}) and in the rest of the paper.}\BibitemShut {Stop}%
\bibitem [{com({\natexlab{a}})}]{comment_fidelity}%
  \BibitemOpen
  \href@noop {} {}\bibinfo {note} {For $\delta R/\delta x \lesssim 1$ the
  approximation of $1-\max[\overline{P_\downarrow (t)} ]$ given in
  Eq.~(\ref{infidelity}) is very close to its upper bound. However, at large
  values of $\delta R$, Eq.~(\ref{infidelity}) gives
  $1-\max[\overline{P_\downarrow (t)} ] \simeq \sqrt{2/\pi^3}(4+\pi^2) \eta ^2
  (1.29+\ln \delta R/\delta x)/(\delta R/\delta x)^3$, which approaches zero
  faster than the upper bound.}\BibitemShut {Stop}%
\bibitem [{sup()}]{supplemental}%
  \BibitemOpen
  \href@noop {} {}\bibinfo {note} {See the Supplemental Information for a more
  detailed analysis of the stationary value.}\BibitemShut {Stop}%
\bibitem [{\citenamefont {Nadj-Perge}\ \emph {et~al.}(2012)\citenamefont
  {Nadj-Perge}, \citenamefont {Pribiag}, \citenamefont {van~den Berg},
  \citenamefont {Zuo}, \citenamefont {Plissard}, \citenamefont {Bakkers},
  \citenamefont {Frolov},\ and\ \citenamefont {Kouwenhoven}}]{Nadj-Perge2012}%
  \BibitemOpen
  \bibfield  {author} {\bibinfo {author} {\bibfnamefont {S.}~\bibnamefont
  {Nadj-Perge}}, \bibinfo {author} {\bibfnamefont {V.~S.}\ \bibnamefont
  {Pribiag}}, \bibinfo {author} {\bibfnamefont {J.~W.~G.}\ \bibnamefont
  {van~den Berg}}, \bibinfo {author} {\bibfnamefont {K.}~\bibnamefont {Zuo}},
  \bibinfo {author} {\bibfnamefont {S.~R.}\ \bibnamefont {Plissard}}, \bibinfo
  {author} {\bibfnamefont {E.~P. A.~M.}\ \bibnamefont {Bakkers}}, \bibinfo
  {author} {\bibfnamefont {S.~M.}\ \bibnamefont {Frolov}}, \ and\ \bibinfo
  {author} {\bibfnamefont {L.~P.}\ \bibnamefont {Kouwenhoven}},\ }\href@noop {}
  {\bibfield  {journal} {\bibinfo  {journal} {Phys. Rev. Lett.}\ }\textbf
  {\bibinfo {volume} {108}},\ \bibinfo {pages} {166801} (\bibinfo {year}
  {2012})}\BibitemShut {NoStop}%
\bibitem [{com({\natexlab{b}})}]{comment_motional_narrowing}%
  \BibitemOpen
  \href@noop {} {}\bibinfo {note} {When $\delta R/\delta x \gg 1$, the
  large-amplitude motion induces an averaging of several independent nuclear
  configurations, separated by a distance $\sim \delta x$ \cite{Palyi2014}.
  This in turn leads to a decrease of $\delta h_{xy}$.}\BibitemShut {Stop}%
\bibitem [{\citenamefont {Petta}\ \emph {et~al.}(2005)\citenamefont {Petta},
  \citenamefont {Johnson}, \citenamefont {Taylor}, \citenamefont {Laird},
  \citenamefont {Yacoby}, \citenamefont {Lukin}, \citenamefont {Marcus},
  \citenamefont {Hanson},\ and\ \citenamefont {Gossard}}]{Petta2005}%
  \BibitemOpen
  \bibfield  {author} {\bibinfo {author} {\bibfnamefont {J.~R.}\ \bibnamefont
  {Petta}}, \bibinfo {author} {\bibfnamefont {A.~C.}\ \bibnamefont {Johnson}},
  \bibinfo {author} {\bibfnamefont {J.~M.}\ \bibnamefont {Taylor}}, \bibinfo
  {author} {\bibfnamefont {E.~A.}\ \bibnamefont {Laird}}, \bibinfo {author}
  {\bibfnamefont {A.}~\bibnamefont {Yacoby}}, \bibinfo {author} {\bibfnamefont
  {M.~D.}\ \bibnamefont {Lukin}}, \bibinfo {author} {\bibfnamefont {C.~M.}\
  \bibnamefont {Marcus}}, \bibinfo {author} {\bibfnamefont {M.~P.}\
  \bibnamefont {Hanson}}, \ and\ \bibinfo {author} {\bibfnamefont {A.~C.}\
  \bibnamefont {Gossard}},\ }\href {\doibase 10.1126/science.1116955}
  {\bibfield  {journal} {\bibinfo  {journal} {Science}\ }\textbf {\bibinfo
  {volume} {309}},\ \bibinfo {pages} {2180} (\bibinfo {year}
  {2005})}\BibitemShut {NoStop}%
\bibitem [{\citenamefont {Kawakami}\ \emph {et~al.}(2014)\citenamefont
  {Kawakami}, \citenamefont {Scarlino}, \citenamefont {Ward}, \citenamefont
  {Braakman}, \citenamefont {Savage}, \citenamefont {Lagally}, \citenamefont
  {Friesen}, \citenamefont {Coppersmith}, \citenamefont {Eriksson},\ and\
  \citenamefont {Vandersypen}}]{Kawakami2014}%
  \BibitemOpen
  \bibfield  {author} {\bibinfo {author} {\bibfnamefont {E.}~\bibnamefont
  {Kawakami}}, \bibinfo {author} {\bibfnamefont {P.}~\bibnamefont {Scarlino}},
  \bibinfo {author} {\bibfnamefont {D.~R.}\ \bibnamefont {Ward}}, \bibinfo
  {author} {\bibfnamefont {F.~R.}\ \bibnamefont {Braakman}}, \bibinfo {author}
  {\bibfnamefont {D.~E.}\ \bibnamefont {Savage}}, \bibinfo {author}
  {\bibfnamefont {M.~G.}\ \bibnamefont {Lagally}}, \bibinfo {author}
  {\bibfnamefont {M.}~\bibnamefont {Friesen}}, \bibinfo {author} {\bibfnamefont
  {S.~N.}\ \bibnamefont {Coppersmith}}, \bibinfo {author} {\bibfnamefont
  {M.~A.}\ \bibnamefont {Eriksson}}, \ and\ \bibinfo {author} {\bibfnamefont
  {L.~M.~K.}\ \bibnamefont {Vandersypen}},\ }\href@noop {} {\bibfield
  {journal} {\bibinfo  {journal} {Nature Nanotechnology}\ }\textbf {\bibinfo
  {volume} {9}},\ \bibinfo {pages} {666} (\bibinfo {year} {2014})}\BibitemShut
  {NoStop}%
\bibitem [{\citenamefont {Veldhorst}\ \emph {et~al.}(2014)\citenamefont
  {Veldhorst}, \citenamefont {Hwang}, \citenamefont {Yang}, \citenamefont
  {Leenstra}, \citenamefont {de~Ronde}, \citenamefont {Dehollain},
  \citenamefont {Muhonen}, \citenamefont {Hudson}, \citenamefont {Itoh},
  \citenamefont {Morello},\ and\ \citenamefont {Dzurak}}]{Veldhorst2014}%
  \BibitemOpen
  \bibfield  {author} {\bibinfo {author} {\bibfnamefont {M.}~\bibnamefont
  {Veldhorst}}, \bibinfo {author} {\bibfnamefont {J.~C.~C.}\ \bibnamefont
  {Hwang}}, \bibinfo {author} {\bibfnamefont {C.~H.}\ \bibnamefont {Yang}},
  \bibinfo {author} {\bibfnamefont {A.~W.}\ \bibnamefont {Leenstra}}, \bibinfo
  {author} {\bibfnamefont {B.}~\bibnamefont {de~Ronde}}, \bibinfo {author}
  {\bibfnamefont {J.~P.}\ \bibnamefont {Dehollain}}, \bibinfo {author}
  {\bibfnamefont {J.~T.}\ \bibnamefont {Muhonen}}, \bibinfo {author}
  {\bibfnamefont {F.~E.}\ \bibnamefont {Hudson}}, \bibinfo {author}
  {\bibfnamefont {K.~M.}\ \bibnamefont {Itoh}}, \bibinfo {author}
  {\bibfnamefont {A.}~\bibnamefont {Morello}}, \ and\ \bibinfo {author}
  {\bibfnamefont {A.~S.}\ \bibnamefont {Dzurak}},\ }\href@noop {} {\bibfield
  {journal} {\bibinfo  {journal} {Nature Nanotechnology}\ }\textbf {\bibinfo
  {volume} {9}},\ \bibinfo {pages} {981} (\bibinfo {year} {2014})}\BibitemShut
  {NoStop}%
\end{thebibliography}%

\newpage

\pagebreak
\widetext
\begin{center}
\textbf{\large Supplemental material for ``Dephasing due to nuclear spins in large-amplitude electric dipole spin resonance"}
\end{center}
\setcounter{equation}{0}
\setcounter{figure}{0}
\setcounter{table}{0}
\setcounter{page}{1}
\makeatletter
\renewcommand{\theequation}{S\arabic{equation}}
\renewcommand{\thefigure}{S\arabic{figure}}
\renewcommand{\bibnumfmt}[1]{[S#1]}
\renewcommand{\citenumfont}[1]{S#1}

\setlength\arraycolsep{0pt}

\begin{figure}[h]
\begin{center}
\includegraphics[height=0.31\textwidth]{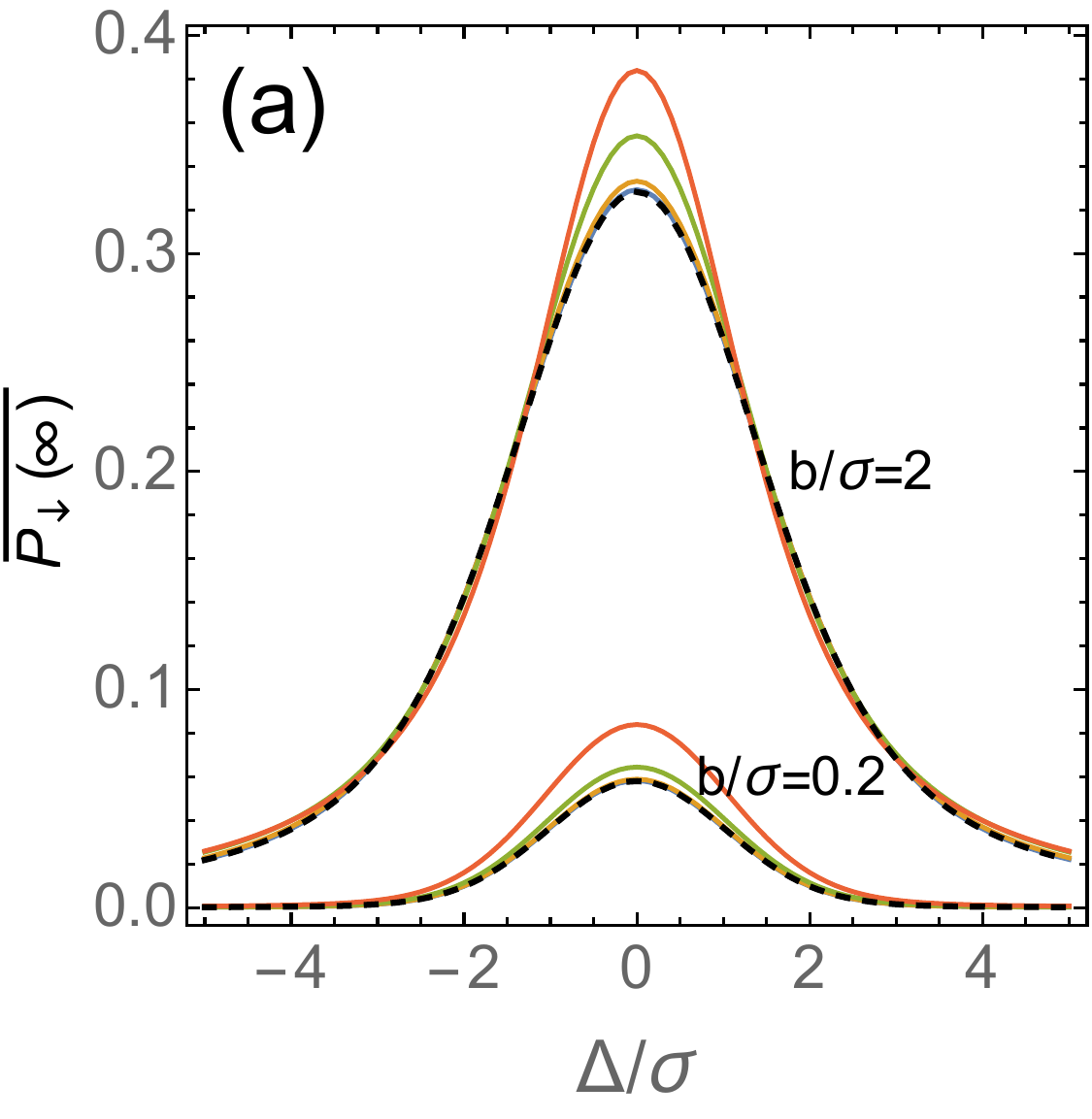}
\includegraphics[height=0.3\textwidth]{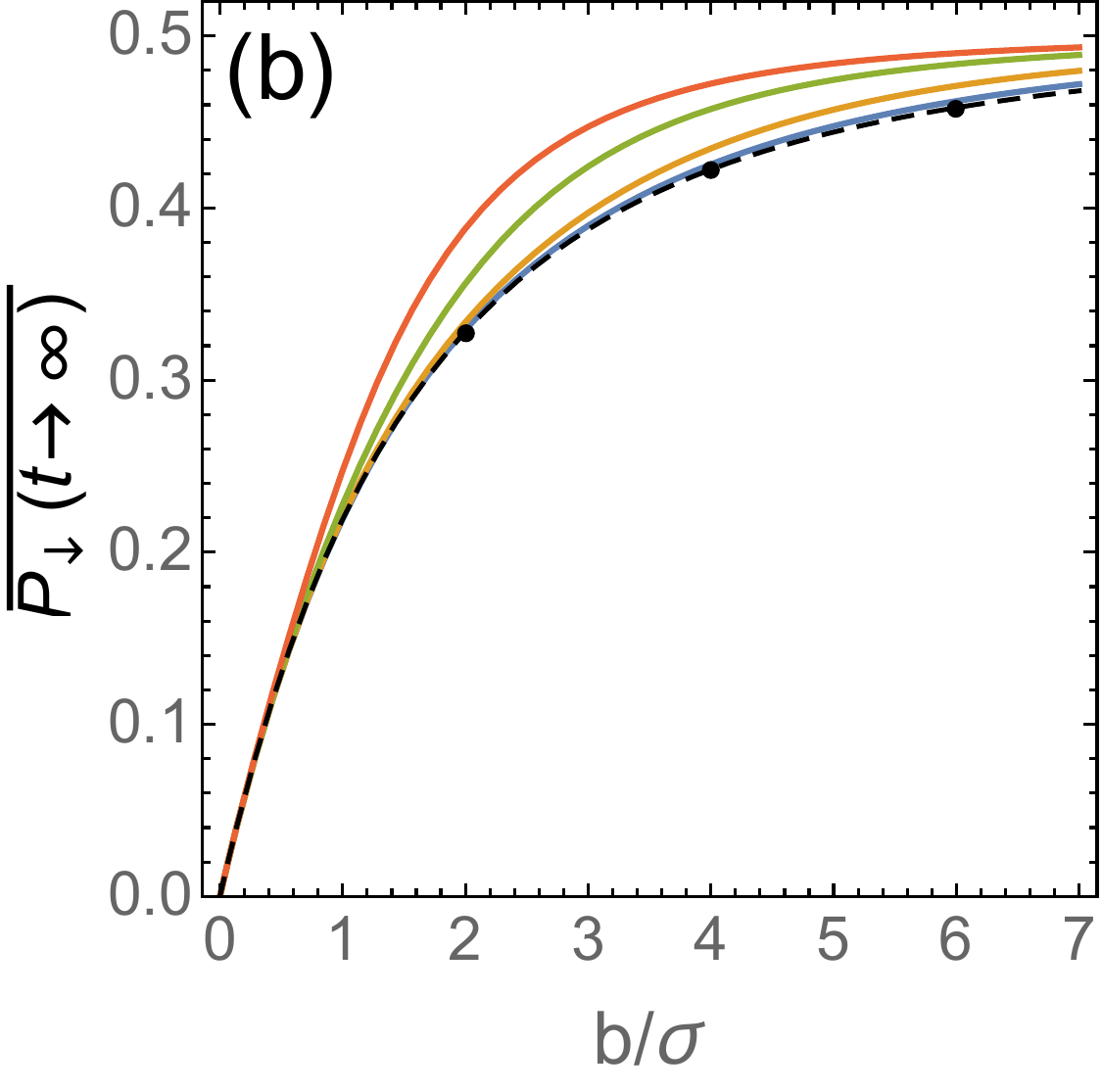}
\includegraphics[height=0.31\textwidth]{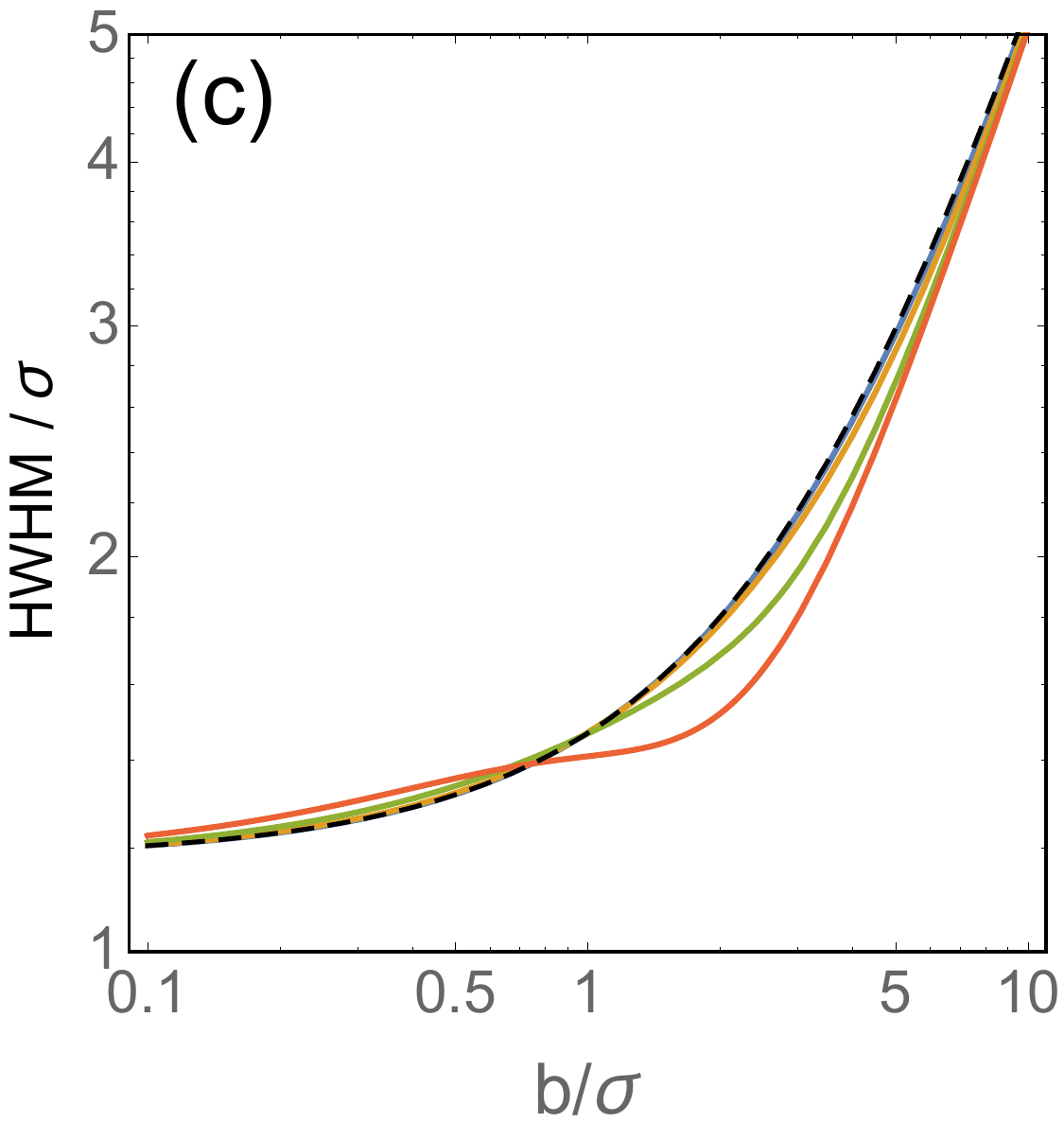}
\caption{
Characteristics of $\overline{P_\downarrow(t \to \infty)}$. Each family of curves shows results with $\eta =0.05,0.1,0.2,0.5,1$. Larger values of $\eta$ correspond to larger deviations from the $\eta=0.05$ results (dashed), which are essentially indistinguishable from $\eta=0$. In panel (a) we plot $\overline{P_\downarrow(\infty)} $ as function of detuning, at two representative drive strengths
$b/\sigma = 0.2$ and 2. Panel (b) is the $\Delta =0$ value. The dots on the dashed curve ($\eta = 0.05$) mark the stationary values for the three lowest curves of Fig.~1(b) of the main text. The peak around $\Delta=0$ is characterized by the half width at half-maximum plotted in panel (c).
\label{fig:linewidth}}
\end{center}
\end{figure}

We discuss here here the behavior of $\overline{P_\downarrow(t \to \infty)} $, i.e., the stationary value of the spin-flip probability under a continuous drive. As a function of detuning, $\overline{P_\downarrow(\infty)} $ displays a peak around $\Delta =0$ and, like in regular ESR (the $\eta \to 0$ limit), a stronger drive leads to a general increase of $\overline{P_\downarrow(\infty)}$, as well as broadening in $\Delta$.  Representative examples are shown if Fig.~\ref{fig:linewidth}(a).

We first examine the behavior of the peak value as a function of the drive strength. Taking $\Delta =0$ in Eq.~(9) of the main text gives $\overline{P_\downarrow(\infty)} \simeq 1/2$ in the large-drive regime, thus the interesting drive dependence of $\overline{P_\downarrow(\infty)}$ occurs when $h_{x,y}$ are smaller than $b,h_z$. By neglecting  in Eq.~(8) such transverse nuclear fluctuations, we obtain the following expression:
\begin{equation}\label{asympt}
\overline{ P_\downarrow(\infty) } \simeq  \frac{\sqrt{\pi}b}{4\sqrt{2}\sigma \delta h_z} e^{b^2/(8 \sigma^2\delta h_z^2)}
{\rm erfc} \left(\frac{b}{2\sqrt{2} \sigma \delta h_z}\right),
\end{equation}
which is similar to the ESR result, except that the nuclear fluctuations $\sigma$ are replaced here by a reduced value $\sigma \delta h_z$. We have checked that Eq.~(\ref{asympt}) is in agreement with direct numerical evaluation. As seen in Fig.~\ref{fig:linewidth}(b), deviations from the ESR result exist in general but in the current experimental regime can be safely neglected, because they only become important when $\eta \sim 1$. 

By considering the dependence of $\overline{P_\downarrow(\infty)}$ on detuning, we can reach a similar conclusions about the EDSR linewidth. Figure~\ref{fig:linewidth}(c) shows that transverse fluctuations have a significant effect for $\eta \sim 1$ and $b \gtrsim \sigma$, when the linewidth is significantly narrower than for ESR.  However, very small deviations from the Voigt profile occur for currently more typical values $\eta \ll 1$.  As stated in the main text, we can thus conclude that $\overline{P_\downarrow(\infty)}$ is much less sensiteve to the physical origin of the drive than the decay of Rabi oscillations. One can see, contrasting Fig.~\ref{fig:linewidth}(b) with Fig.~1(b) of the main text, that significant deviations from the ESR decay appear at relatively small values of $\delta R/\delta x$, when $\overline{P_\downarrow(\infty)}$ has not yet saturated to 1/2. 

\end{document}